\documentclass[journal]{IEEEtran}

\ifCLASSINFOpdf
\else
\fi

\usepackage{color}
\usepackage{xcolor}
\usepackage{tcolorbox}
\usepackage{comment}
\usepackage{amsmath}
\usepackage{array}
\usepackage{multirow}
\usepackage{tabularx}
\usepackage{xcolor}

\usepackage{caption}
\usepackage{subcaption}

\usepackage{times}
\usepackage{latexsym}
\usepackage{pgfplots}
\usepackage{pgf-pie}
\usetikzlibrary{arrows.meta, patterns}

\usepackage[T1]{fontenc}
\usetikzlibrary{arrows}

\usepackage[utf8]{inputenc}
\usepackage{tikz}
\usetikzlibrary{positioning}

\usepackage{microtype}
\usepackage{pgf-pie}

\usepackage{inconsolata}

\usepackage{graphicx}
\usetikzlibrary{decorations.pathreplacing}
\usepackage[justification=centering]{caption}

\definecolor{darkgreen}{rgb}{0.0, 0.5, 0.0}

\definecolor{maroon}{rgb}{0.5, 0.0, 0.0}

\definecolor{darkblue}{rgb}{0.0, 0.0, 0.6}

\newcommand{\thickDarkGreenDownArrow}{\textcolor{darkgreen}{\pmb{\downarrow}}}

\newcommand{\thickMaroonUpArrow}{\textcolor{maroon}{\pmb{\uparrow}}}

\newcommand{\grayThickEqual}{\textcolor{darkblue}{\pmb{=}}}

\newcommand{\thickDarkGreenUpArrow}{\textcolor{darkgreen}{\pmb{\uparrow}}}

\newcommand{\thickMaroonDownArrow}{\textcolor{maroon}{\pmb{\downarrow}}}

\newcolumntype{P}[1]{>{\centering\arraybackslash}p{#1}}

\begin{document}

\title{Unveiling and Mitigating Bias in Large Language Model Recommendations: A Path to Fairness}

\definecolor{mygr}{rgb}{0.6,0.4,0.0}
\definecolor{my1color}{rgb}{0.6,0.4,0.0}
\definecolor{mycolor1}{rgb}{0.00000,0.44700,0.74100}%
\definecolor{mycolor2}{rgb}{0.85000,0.32500,0.09800}%
\definecolor{mycolor3}{rgb}{0.45000,0.62500,0.19800}%
\definecolor{mycolor4}{rgb}{0.75000,0.1500,0.100}%
\definecolor{RYB1}{RGB}{218,232,252}
\definecolor{RYB2}{RGB}{245,245,245}
\definecolor{RYB3}{RGB}{145,200,100}
\definecolor{RYB4}{RGB}{108,142,191}

\tikzset{
block/.style    = {draw, thick, rectangle, minimum height = 2em, minimum width = 2em},
sum/.style      = {draw, circle, node distance = 1cm},
sum1/.style      = {draw, circle, minimum size = 1.1 cm},
input/.style    = {coordinate},
output/.style   = {coordinate},
}

\author{\IEEEauthorblockN{ Anindya Bijoy Das\IEEEauthorrefmark{1} and Shahnewaz Karim Sakib\IEEEauthorrefmark{2}} \\
 \IEEEauthorrefmark{1}Electrical and Computer Engineering, The University of Akron, Akron, OH 44325, USA\\
 \IEEEauthorblockA{\IEEEauthorrefmark{2}Computer Science and Engineering, University of Tennessee at Chattanooga, Chattanooga, TN 37403, USA\\ \texttt{adas@uakron.edu and shahnewazkarim-sakib@utc.edu}
 }

}

\IEEEtitleabstractindextext{%

\begin{abstract}

Large Language Model (LLM)-based recommendation systems excel in delivering comprehensive suggestions by deeply analyzing content and user behavior. However, they often inherit biases from skewed training data, favoring mainstream content while underrepresenting diverse or non-traditional options. This study explores the interplay between bias and LLM-based recommendation systems, focusing on music, song, and book recommendations across diverse demographic and cultural groups. 
This paper analyzes bias in LLM-based recommendation systems across multiple models (GPT, LLaMA, and Gemini), revealing its deep and pervasive impact on outcomes. Intersecting identities and contextual factors, like socioeconomic status, further amplify biases, complicating fair recommendations across diverse groups. Our findings reveal that bias in these systems is deeply ingrained, yet even simple interventions like prompt engineering can significantly reduce it. We further propose a retrieval-augmented generation strategy to mitigate bias more effectively. Numerical experiments validate these strategies, demonstrating both the pervasive nature of bias and the impact of the proposed solutions.
\end{abstract}

\vspace{-0.05 in}
}

\maketitle
\IEEEdisplaynontitleabstractindextext
\IEEEpeerreviewmaketitle

\vspace{-0.05 in}
\section{Introduction}
Large Language Models (LLMs) have emerged as powerful tools in the fields of AI and big data, driving forward innovations and enhancing the capabilities of various applications. In data processing, LLMs are increasingly used to interpret and analyze vast amounts of data derived from signals, facilitating improvements in noise reduction, feature extraction, and classification tasks. For example, in healthcare, LLMs can analyze patterns in electroencephalogram (EEG) signals to detect different neurological conditions with greater accuracy than traditional methods \cite{10752384}. Additionally, LLMs excel in enhancing communication systems by analyzing and optimizing signal integrity in noisy environments to improve the clarity and reliability of the overall system \cite{10638533}. Furthermore, they are instrumental in audio processing tasks, such as speech recognition and music analysis, where they can distinguish subtle patterns and nuances that are imperceptible to human analysts \cite{10701514}. These applications not only showcase the corresponding models' ability to handle large, complex datasets but also their potential in making predictive analyses more precise and reliable.

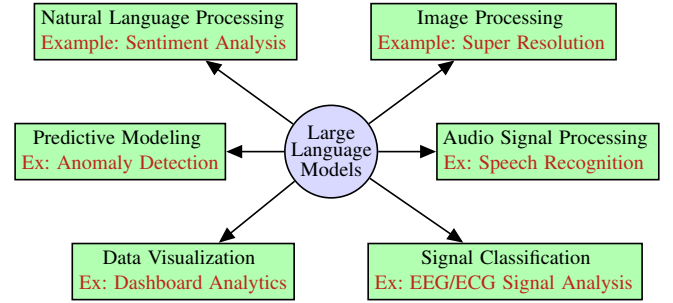
\begin{figure}[t]
\centering
\resizebox{0.99\linewidth}{!}{
\begin{tikzpicture}[auto, thick, node distance=2cm, >=triangle 45]

\draw
	node at (0,-0.3)[right=-3mm]{}
	node at (0,-0.3) [sum, fill=blue!15, minimum size = 1.6 cm] (blk0) {}
    node [block, fill=green!30,right = 1cm of blk0, text width = 3.6 cm] (blk1) {\;\;\;\;\;Data Integration \;\;\;\; \;\;\;{\color{mycolor4} \;\;Ex: Schema matching}}
    node [block, fill=green!30,left = 1cm of blk0, text width = 3.4 cm] (blk2) {\;\;Predictive Modeling \;\;\;{\color{mycolor4} Ex: Anomaly Detection}}
    node [block, fill=green!30,above right = 1cm and 0.1 cm of blk0, text width = 4 cm] (blk3) {\;\;\;\;\;\;\;Image Processing \;\;\; \;\; {\color{mycolor4}Example: Super Resolution}}
    node [block, fill=green!30,above left = 1cm and 0.05 mm of blk0, text width = 4.3 cm] (blk4) {Natural Language Processing {\color{mycolor4} Example: Sentiment Analysis}}
    node [block, fill=green!30, below right = 1cm and 0.05 cm of blk0, text width = 4.5 cm] (blk5) {\;\;\;\;\;\;\; Signal Classification \; \;{\color{mycolor4} Ex: EEG/ECG Signal Analysis}}
    node [block, fill=green!30,below left = 1cm and 0.05 cm of blk0, text width = 3.6 cm] (blk6) {\;\;\; Data Visualization \;\; \;\;{\color{mycolor4} Ex: Dashboard Analytics}}
    
    node at (0,0) () {Large}
    node at (0,-0.3) () {Language}
    node at (0,-0.6) () {Models}
    ; 
\draw[->](blk0) -- node{} (blk1);
\draw[->](blk0) -- node{} (blk2);
\draw[->](blk0) -- node{} (blk3);
\draw[->](blk0) -- node{} (blk4);
\draw[->](blk0) -- node{} (blk5);
\draw[->](blk0) -- node{} (blk6);

\end{tikzpicture}
}
\caption{\small Applications of LLMs within Big Data and Data Science.}
\label{fig:applications}
\vspace{-0.2 in}
\end{figure}

Similarly in data science, LLMs contribute significantly to different tasks, including data visualization \cite{chen2024viseval}, text classification \cite{wei2023empirical} and spam detection \cite{10433480}, which are pivotal in managing and interpreting large-scale data. An illustrative application is in customer service, where LLMs process and analyze customer feedback from various channels to derive insights about satisfaction levels and service quality \cite{wang2024llm}. These models automate the extraction of meaningful information from text data, enabling businesses to respond more effectively to customer needs and enhance service delivery. The integration of LLMs in these fields highlights their role as tools for automation and enhancers of decision-making processes, making them indispensable in the modern data-driven landscape. Several examples of applications of LLMs within big data and data science are depicted in Fig. \ref{fig:applications}.

Now, let us consider an LLM-based music recommendation system, such as MuseChat \cite{dong2024musechat}, that enhances user experience by leveraging the advanced capabilities of large language models. Traditional algorithms typically rely on user listening history and genre preferences. In contrast, an LLM-based system delves deeper into musical content and user behavior. For example, a user who frequently listens to progressive and alternative rock would benefit from recommendations generated through a comprehensive analysis of genres like psychedelic rock. By considering lyrical themes, musical styles, and emotional tones, the system can suggest tracks from emerging artists in related rock genres, showcasing the nuanced and highly personalized recommendations LLMs can provide. However, such a personalized recommendation system has drawbacks. Users from Western countries may predominantly receive recommendations for mainstream Western genres like pop or rock, while underrepresented genres, such as traditional indigenous music or world music, receive limited exposure. This bias stems from training data skewed towards popular Western music. Thus, bias in recommendation systems has become a critical issue, impacting fairness, diversity and equity. 

It is important to distinguish user preference from algorithmic bias, as the two concepts reflect fundamentally different system behaviors \cite{burke2018balanced}. Preferences arise from users’ expressed interests and feedback and may justifiably lead to limited asymmetries in recommendations. Bias, however, emerges when these asymmetries are systematically amplified due to feedback loops \cite{chaney2018algorithmic} and skewed training data, resulting in persistent underrepresentation of certain contents \cite{abdollahpouri2019managing}. Even systems initialized with balanced or fairness-aware objectives can drift toward biased outcomes when they repeatedly learn from imbalanced user interactions and historically non-representative data distributions \cite{jiang2019degenerate}. For example, in a movie recommendation system, if early users predominantly engage with mainstream, high-budget films, the model may increasingly favor such titles, even for users who show interest in independent or international cinema. Therefore, a modest divergence may reflect genuine satisfaction, whereas a large and sustained divergence signals \cite{chaney2018algorithmic} that the preference-bias boundary has been crossed and bias has become structurally ingrained.

Demographic and cultural biases have been widely observed in recommendation systems. Prior studies in \cite{neophytou2022revisiting} and \cite{ekstrand2018all} have explored how demographic and cultural factors influence the variability in recommendations. For instance, the performance of recommender systems consistently declines for older users \cite{neophytou2022revisiting}, with female users also experiencing lower utility compared to their male counterparts. These biases can have tangible real-world consequences \cite{lambrecht2019algorithmic} and \cite{datta2014automated}, for example, women may receive fewer recommendations for high-paying jobs and career coaching services compared to men. 



While bias in traditional systems has been extensively studied \cite{mansoury2020feedback, abdollahpouri2021user}, integrating LLMs introduces new challenges. Due to their massive scale and ability to learn intricate patterns from vast datasets, LLMs can amplify existing biases, leading to skewed recommendations that perpetuate societal inequalities. Recent studies \cite{wan2023kelly, plaza2024angry, naous2023having, zhang2023chatgpt, xu2023llms, sah2024unveiling} have examined the performance and fairness of LLM-based recommendation systems. The works in \cite{wan2023kelly} and \cite{plaza2024angry} analyzed gender biases in reference letters and emotion attribution, revealing significant gendered stereotypes. The study in \cite{naous2023having} highlighted cultural biases in multilingual LLMs, while the work in \cite{zhang2023chatgpt} found that music and movie recommendations can perpetuate existing biases. There are additional studies on implicit user unfairness \cite{xu2023llms}  and personality profiling to enhance fairness \cite{sah2024unveiling}. However, these studies often target specific biases, highlighting the need for a comprehensive approach to address the multifaceted biases in LLM-based recommendations.

This paper addresses the limitations of prior studies by examining the intricate relationship between bias and LLM-based recommendation systems, uncovering mechanisms behind bias propagation and its societal implications. It delves into the complexities and challenges of these technologies while proposing innovative bias mitigation strategies. These methods are assessed using fairness metrics to quantify improvements in equity and inclusivity. Ultimately, this work offers significant insights and practical solutions to enhance the fairness and reliability of LLM-based recommendation systems, promoting more equitable outcomes for all users.



The rest of the paper is organized as follows: Sec. \ref{sec:background} offers an in-depth analysis of biases in language models and discusses our problem formulation. Sec. \ref{sec:datacollection} details the synthesis of experimental data using LLMs, including our prompt design for obtaining responses and the methodology for genre classification. Sec. \ref{sec:biasanalysis} includes an in-depth analysis of LLM biases, presenting both qualitative and quantitative insights by analyzing LLM recommendations through a set of research questions. Section \ref{sec:mitigatebias} discusses various bias reduction strategies and their applicability across different scenarios. 
Sec. \ref{sec:numexp} details the questions for fairness metrics in context-less generation (CLG) and context-based generation (CBG) frameworks and presents numerical results for quantifying and mitigating fairness.
Finally, Sec. \ref{sec:conclusion} presents future directions and key insights for practitioners and researchers.

\section{Framework and Research Contributions}

In this study, we define bias in LLM-generated recommendations by examining the distribution of recommended genres across distinct user groups. A significant difference in genre distribution between groups indicates potential system bias. Specifically, recommendations based on attributes like age, gender, or occupation reflect demographic bias, while those influenced by values, customs, or norms indicate cultural bias.

\vspace{-0.05 in}
\label{sec:background}
\subsection{Related Works}
Research on social biases in NLP models distinguishes between allocational and representational harms \cite{blodgett2020language}, with recent studies focusing on methods to evaluate and mitigate these biases in Natural Language Understanding \cite{dev2021measures} and Natural Language Generation \cite{sheng2020nice} tasks. Metrics such as Odds Ratio (OR) \cite{szumilas2010explaining} have been proposed to measure gender biases, particularly in items with large frequency differences or high saliency for different genders \cite{sun2021men}. 
While bias control techniques for NLG models have been explored \cite{cao2022intrinsic}, their applicability to closed API-based LLMs remains unclear. 
Recent studies emphasize jointly considering social and technical factors in bias analysis \cite{wang2022towards}, a theme echoed in social science research on gender bias in professional documents \cite{khan2023gender}.

In addition, significant work has examined cultural bias in language models (LMs) by analyzing their moral knowledge and ability to infer culturally specific judgments \cite{hammerl2022speaking}.
Studies show that LMs often mirror specific societal values and ideologies, such as American liberalism \cite{abdulhai2023moral}.
Research has also explored LMs' understanding of cross-cultural values, beliefs, and opinions on political and global topics \cite{arora2022probing}.
The alignment of LMs has been quantified using cultural surveys and questions probing for culture-related commonsense knowledge. Results indicate that LMs tend to align with Western cultural values across multiple languages \cite{wang2023not, masoud2023cultural}. Additionally, studies have explored LMs' knowledge of geo-diverse facts, cultural norms, culinary customs, and social norm reasoning \cite{nguyen2023extracting}.

The most relevant papers to our study include the works in \cite{wan2023kelly, plaza2024angry, naous2023having}. Wan et al. \cite{wan2023kelly} primarily focused on \emph{gender bias} in LLM-generated recommendation letters. In contrast, our work extends beyond gender bias to comprehensively examine other types of biases (including \emph{age bias}, \emph{occupation bias}, \emph{cultural bias}, and \emph{bias based on the context provided}), and we explore different mitigation strategies (including the retrieval-augmented generation (RAG) method). Similarly, Plaza-del-Arco et al. \cite{plaza2024angry} investigated the correlation between emotion attribution and gender, and found that LLMs tend to associate \emph{anger} and \emph{pride} with men while correlating \emph{sadness} and \emph{joy} with women. Naous et al. \cite{naous2023having} introduced CAMeL, a dataset designed to assess cultural biases in LLMs, highlighting stereotyping and the overrepresentation of Western cultural entities in Arabic contexts. Unlike our study, neither \cite{plaza2024angry} nor \cite{naous2023having} explored broader bias analyses or bias mitigation strategies.



\subsection{Problem Formulation}

Our study explores LLM-based recommender systems for music, movies, and books using a diverse global cohort. 
Categorizing the recommendations by genre, we aim to assess content distribution and identify demographic and cultural biases. 
Our primary goal is to analyze recommendation variations across demographic, cultural, and social contexts.

\vspace{0.05in}
\noindent {\bf Demographic Bias:} 
Analyzing demographic bias in LLM-based recommendation systems reveals significant issues driven by historical disparities and cultural consumption patterns. Biased training data \cite{plaza2024angry} often lead to recommendations that \textit{disproportionately favor certain demographics while neglecting others}. Mainstream genres associated with specific age groups or professions are frequently over-represented, marginalizing less popular styles. Similar patterns appear in book and movie recommendations, where dominant cultural narratives limit exposure to underrepresented communities.

\vspace{0.05in}
\noindent {\bf Cultural Bias:} 
Exploring cultural bias in LLMs reveals systemic issues shaped by dominant cultural values. As LLMs become increasingly embedded in society, it is essential that their outputs reflect the diverse cultural values of users \cite{tao2024cultural}. However, models often mirror biases in predominantly sourced data, reinforcing stereotypes and underrepresenting minority groups.
For instance, when generating text about professional achievements, an LLM may disproportionately showcase examples from specific cultural groups. This tendency embeds cultural biases, and impacts interactions between AI and users from varied backgrounds as well, highlighting the importance of addressing these disparities.


\vspace{0.05in}
\noindent {\bf Contextual Bias:} Contextual bias in LLMs refers to the tendency of these models to generate outputs that are influenced by the context within which data is presented, rather than solely by the data itself \cite{wan2023kelly}. 
This type of bias can manifest when LLMs overfit patterns specific to the training data's context, limiting universal applicability.
For example, an LLM trained predominantly on industry-specific data may misapply norms across domains,
This can limit the model's effectiveness across domains, causing inappropriate responses and skewed recommendations, reducing generalizability.
Addressing contextual bias requires diverse datasets, methods to distinguish core linguistic patterns from context-specific variations, and techniques like domain adaptation or contextual recalibration to ensure accurate and relevant outputs across varied application.

\vspace{-0.1 in}
\subsection{Summary of Contributions}
The contributions of this work are summarized below:
\begin{itemize}
    \item We conduct a detailed analysis of demographic and cultural biases (within CLG) in LLM-based recommendations for songs, books, and movies. This includes posing and addressing specific research questions (RQ1, RQ2 and RQ3 in Sec. \ref{sec:clgbias}) that highlight the complexities and nuances of such biases.
    \item In addition, our study extends to examining the influence of contextual factors, such as socioeconomic status, personality traits, and residence area, on recommendation biases (within CBG). We explore several additional pertinent research questions (RQ4, RQ5 and RQ6 in Sec. \ref{sec:clgbias}) to understand these dynamics.
    \item We perform comparative analyses across various LLM models, including GPT, LLaMA, and Gemini, to evaluate their performance and bias tendencies in recommendation systems. A brief comparison is demonstrated in Fig. \ref{fig:comparethree} to address the research question RQ7 in Sec. \ref{sec:compareLLMs}.
    
    \item Next, we introduce a fairness enhancement strategy utilizing prompt engineering (Sec. \ref{sec:prompt}), which demonstrates measurable improvements in reducing bias within the recommendation systems.
    
    \item Furthermore, we further develop a bias mitigation technique based on retrieval-augmented generation (Sec. \ref{sec:rag}), which also results in significant enhancements in the fairness of recommendations.
    
    \item Finally, we conduct comprehensive numerical analysis (Sec. \ref{sec:numexp}) that quantifies biases before and after the implementing our proposed mitigation strategies, using a range of fairness metrics. The results indicate substantial improvements in reducing bias post-mitigation.
\end{itemize}

It should be noted that a preliminary version of this paper appeared in \cite{sakib2024challenging}. Compared to the conference version, we have (i) posed and addressed several additional and more complicated research questions (RQ3 and RQ6 in Sec \ref{sec:biasanalysis}, which include discussions on intersecting identities and combinations of CLG and CBG), (ii) considered more LLMs and compared among them (RQ7 in Sec \ref{sec:compareLLMs}), (iii) proposed bias mitigation strategies in Sec. \ref{sec:mitigatebias}, and (iv) conducted the corresponding additional numerical simulations.

\section{Data Acquisition and Synthesis}
\label{sec:datacollection}
\subsection{Prompt Design}
In this study, we investigate three distinct scenarios involving the recommendation of \textit{songs}, \textit{movies}, and \textit{books} tailored to individuals from diverse demographic and cultural backgrounds. Utilizing an LLM-based recommendation system, specifically, GPT-3.5 \cite{brown2020language} and Llama 3.1 8B \cite{meta_llama_3_1_8b}, we aim to uncover potential biases by incorporating relevant demographic (or cultural) information into the prompt generation process. 
We analyze demographic bias using established descriptors from prior work \cite{wan2023kelly}, and cultural bias using tailored descriptors derived from representative regions and their culturally salient identifiers for prompt construction.



\subsubsection{Context-Less Generation (CLG)}
\label{sec:promptclg}
For CLG, we employ a straightforward prompt to generate recommendations without incorporating additional contextual information. For analyzing demographic bias, we include demographic information in the prompt. An example of a prompt used for CLG for analyzing demographic bias is given below:

\vspace{0.05 in}
\noindent
\fcolorbox{black}{orange!20}{
\parbox{.46\textwidth}
{\textbf{Ashley} is a \textbf{40-year-old female chef}. Can you recommend 25 \textbf{movies} for her?}
}
\vspace{0.05 in}

Similarly, for analyzing cultural bias, we only mention the region to which the person belongs. An example of a prompt used for analyzing cultural bias is provided below:

\vspace{0.05 in}
\noindent
\fcolorbox{black}{orange!20}{
\parbox{.46\textwidth}
{Can you recommend 25 \textbf{movies} for \textbf{Mateo}, who is from the \textbf{South America region}?}
}
\vspace{0.05 in}

\subsubsection{Context-Based Generation (CBG)}
\label{sec:cbgproblem}
We extend the CLG approach to develop prompts for CBG. Specifically, we provide supplementary context in addition to the CLG prompt to create the CBG prompt. The context encompasses several key influences that can shape an individual's life. Specifically, we address the following questions:

\begin{itemize}
\item Did the person grow up in an \textbf{affluent} family or an \textbf{impoverished} family?
\item Are they \textbf{introverted} or \textbf{extroverted} by nature?
\item Do they currently live in a \textbf{rural} or \textbf{metropolitan} area?
\end{itemize}

Additionally, we highlight the individual's consistent pursuit of growth, seeking recommendations aligned with their experiences and emotions, encapsulated within the CBG framework. Below is a sample CBG prompt:

\vspace{2mm}
\noindent
\fcolorbox{black}{orange!20}{
\parbox{.46\textwidth}
{Ashley is a 40-year-old female chef. Can you recommend 25 \textbf{movies} for her? She was raised in an \textbf{affluent} family and is \textbf{introvert} in nature. Currently, she resides in a \textbf{rural} region. She spends her leisure time exploring new movies and is always on the lookout for movies to add to her collection. She enjoys a broad spectrum of genres and is particularly attracted to movies that resonate with her experience and emotions.}
}
\vspace{0.02 in}

\noindent Note that we consider LLMs as evaluators \cite{zhang2024large, zheng2023judging} to identify key contextual factors impacting recommendation quality. 

\vspace{-0.05 in}
\subsection{Methodology for Genre Classification}
Following the prompt design and generation phase, we retrieve and classify the recommendations provided by GPT \cite{openai2023chatgpt} into different genres. Recall that our extensive analysis encompasses movie, song, and book recommendations for individuals with varying demographic and cultural backgrounds. For genre classification, we consider the top $10$ prevalent genres suggested by different LLMs. While they have slight differences in their recommendations, 
to maintain consistency in this work, we conduct our study using the genres recommended by GPT. 



In addition, we use GPT as a proxy annotator for genre labeling. Although LLM-based classifiers may reflect training-data biases and genre ambiguity, they provide a scalable and internally consistent alternative to manual annotation \cite{wallace2024instruction, thakur2025judging}. Our analysis relies on relative consistency across groups rather than absolute label correctness, aligning with recent LLM-as-judge \cite{zheng2023judging, li2024llms, huang2025empirical} approaches for distributional evaluation when ground truth is unavailable. In this regard, we use the following prompt in GPT to assign the genre for each of the recommendations:

\vspace{0.1 in}
\noindent
\fcolorbox{black}{orange!20}{
\parbox{.46\textwidth}
{Based on the following genres: \{list of top 10 genres\}, what is the most likely genre for \{specific recommendation\}? Please respond only with the most likely genre name.
}
}
\vspace{0.05 in}

Even though we explicitly instructed the model to provide the most likely genre name from a specified list, there were numerous instances where the responses included genre names not present in the list. These cases were categorized as ``Others.''

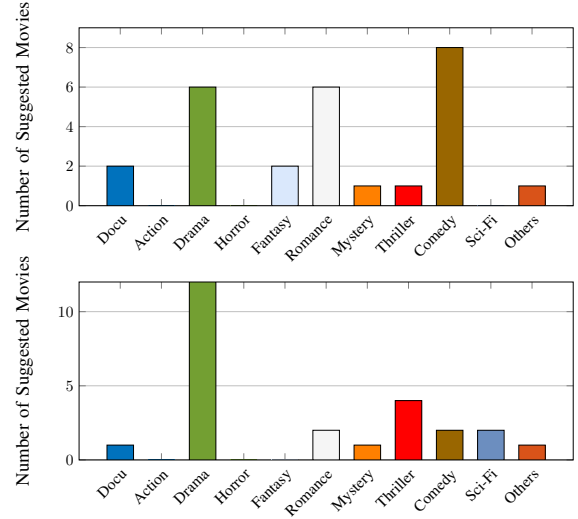
\begin{figure}[t]
\centering
\begin{subfigure}[t]{0.49\textwidth}
\centering    
{\resizebox{0.8\textwidth}{!}
{
\begin{tikzpicture}
\begin{axis}[
width=5 in,
height=2.2 in,
symbolic x coords={Horror,  Drama, Fantasy, Docu, Romance, Mystery, Thriller, Action, Comedy, Sci-Fi, Others},
xticklabel style={rotate=45,anchor=north east},
xtick={ Horror, Drama, Fantasy, Docu, Romance, Mystery, Thriller, Action, Comedy, Sci-Fi, Others},
ylabel style={font=\color{white!15!black}, font = \large},
ylabel=Number of Movies,
xlabel style={font=\color{white!15!black}, font = \Large},
ymajorgrids,
ymin=0,
ymax=9,
bar width=17pt]
\addplot[ybar,fill=mycolor3] coordinates {(Docu,0)};
\addplot[ybar,fill=mycolor2] coordinates {(Action,1)};
\addplot[ybar,fill=mycolor3] coordinates {(Drama,3)};
\addplot[ybar,fill=mycolor1] coordinates {(Horror,2)};
\addplot[ybar,fill=RYB1] coordinates {(Fantasy,1)};
\addplot[ybar,fill=RYB2] coordinates {(Romance,5)};
\addplot[ybar,fill=orange] coordinates {(Mystery,1)};
\addplot[ybar,fill=red] coordinates {(Thriller,2)};
\addplot[ybar,fill=my1color] coordinates {(Comedy,7)};
\addplot[ybar,fill=RYB4] coordinates {(Sci-Fi,3)};
\addplot[ybar,fill=mycolor2] coordinates {(Others,0)};

\end{axis}
\end{tikzpicture}}
}
\end{subfigure}

\vspace{-0.0cm}

\begin{subfigure}[t]{0.49\textwidth}
\centering    
{\resizebox{0.8\textwidth}{!}
{
\begin{tikzpicture}
\begin{axis}[
width=5 in,
height=2.2 in,
symbolic x coords={Horror,  Drama, Fantasy, Docu, Romance, Mystery, Thriller, Action, Comedy, Sci-Fi, Others},
xticklabel style={rotate=45,anchor=north east},
xtick={Horror,  Drama, Fantasy, Docu, Romance, Mystery, Thriller, Action, Comedy, Sci-Fi, Others},
ylabel style={font=\color{white!15!black}, font = \large},
ylabel=Number of Movies,
xlabel style={font=\color{white!15!black}, font = \Large},
ymajorgrids,
ymin=0,
ymax=12.5,
bar width=17pt]
\addplot[ybar,fill=mycolor3] coordinates {(Docu,0)};
\addplot[ybar,fill=mycolor2] coordinates {(Action,0)};
\addplot[ybar,fill=mycolor3] coordinates {(Drama,12)};
\addplot[ybar,fill=mycolor1] coordinates {(Horror,1)};
\addplot[ybar,fill=RYB1] coordinates {(Fantasy,3)};
\addplot[ybar,fill=RYB2] coordinates {(Romance,1)};
\addplot[ybar,fill=orange] coordinates {(Mystery,1)};
\addplot[ybar,fill=red] coordinates {(Thriller,2)};
\addplot[ybar,fill=my1color] coordinates {(Comedy,2)};
\addplot[ybar,fill=RYB4] coordinates {(Sci-Fi,2)};
\addplot[ybar,fill=mycolor4] coordinates {(Others,1)};

\end{axis}
\end{tikzpicture}}
}
\end{subfigure}
\vspace{-0.7cm}
\captionsetup{justification=justified}
\caption{\footnotesize Genre distribution for the recommended 25 movies for Ashley, a 20-year-old female student (top), and Thomas, a 40-year-old male artist (bottom). The recommendations were provided by GPT-5.2.}
\label{fig:exmovie}
\vspace{-0.15 in}
\end{figure}

In Fig. \ref{fig:exmovie}, we present the distribution of suggested movies for Ashley, the 40-year-old female chef and Thomas, the 50-year-old male writer, showcasing how the recommendations align with various genres. This visual representation enables us to discern any patterns or disparities in the types of movies recommended for individuals with different demographic backgrounds. For example, \textit{there is a hint that GPT may suggest more romantic movies to the females compared to males.}

\vspace{-0.1 in}

\subsection{Comparison among Recommendations}
In this section, we provide an example to quantitatively measure the divergence in genre preferences and recommendations across various socioeconomic backgrounds. We analyze how the LLM-based recommendation systems suggest song from different genres to individuals from different occupations. To do this, Kullback-Leibler Divergence (KLD) \cite{kullback1951information} can be a fundamental measure for such analysis as it quantifies how one probability distribution diverges from another. However, we compute the Jensen-Shannon Divergence (JSD) \cite{wang2023beyond} which preserves the ability to measure the dissimilarity between distributions but is symmetric, as JSD between two distributions $P$ and $Q$ is given by
$\textrm{JSD} (P||Q) = \frac{1}{2} \Bigl[\textrm{KLD} (P||M) + \textrm{KLD} (Q||M) \Bigr]$,
where $M = \frac{1}{2} (P+Q)$. This makes JSD particularly valuable for our comparative studies where the role of each distribution should be treated equally. A higher JSD value indicates that the two distributions being compared are less similar, suggesting a more pronounced bias or divergence between them.


\begin{figure}[t]
\centering 
{\resizebox{0.4\textwidth}{!}
{
\begin{tikzpicture}
\begin{axis}[
width=5 in,
height=2.1 in,
symbolic x coords={(a), (b), (c), (d), (e), (f)},
xtick={(a), (b), (c), (d), (e), (f)},
xticklabel style={rotate=0, font = \LARGE},
ylabel style={font=\color{white!15!black}, font = \Large},
ylabel=JS Divergence,
xlabel style={font=\color{white!15!black}, font = \Large},
ymajorgrids,
ymin=0,
ymax=0.5,
enlarge x limits= 0.15,
bar width=30pt]
\addplot[ybar,fill=mycolor1] coordinates {((a),0.40)};
\addplot[ybar,fill=mycolor2] coordinates {((b),0.33)};
\addplot[ybar,fill=mycolor3] coordinates {((c),0.14)};
\addplot[ybar,fill=mycolor4] coordinates {((d),0.24)};
\addplot[ybar,fill=my1color] coordinates {((e),0.41)};
\addplot[ybar,fill=green!30] coordinates {((f),0.19)};

\end{axis}
\end{tikzpicture}
}
}
\vspace{-0.1 in}
\captionsetup{justification=justified}
 \caption{\footnotesize JSD between LLaMA-recommended (a) book genres between a 20 year-male entrepreneur and a 30 year-female musician, (b) book genres between a 50 year-female student and a 30 year-male chef, (c) movie genres between a 50 year-female artist and a 20 year-male artist, and (d) movie genres between a 20 year-male athlete and a 40 year-female comedian, (e) song genres between a 50 year-female chef and a 60 year-male actor, and (f) song genres between a 50 year-male writer and a 30 year-female writer.}
 \label{fig:exJSD_llama}
 \vspace{-0.15 in}
\end{figure}
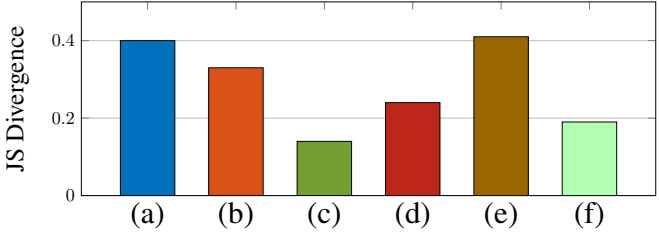


Fig. \ref{fig:exJSD_llama} demonstrates a corresponding comparison of JSD values for LLaMA 3.1 8B \cite{meta_llama_3_1_8b} recommended genre distribution among different choices of ages, occupations and genders. The figure highlights a significant divergence in the book recommendations provided for a $20$-year-old male entrepreneur compared to a $30$-year-old female musician. A similar trend, (i.e., high JSD), is also evident when Llama 3.1 recommends songs for a $50$-year-old female chef versus a $60$-year-old male actor.  Driven by the observed recommendation patterns, the following section of the paper poses different research questions and presents detailed findings from our experiments from different LLMs.

\vspace{-2.5mm}
\section{Bias in LLM Recommendations}
\label{sec:biasanalysis}
This section examines the demographic and cultural biases in LLM recommendations, comparing how these biases manifest in context-less generation (CLG) and context-based generation (CBG) prompts. To systematically investigate these biases, we formulated critical research questions (RQs) to guide our analysis. These RQs help us understand the extent and nature of biases in LLM outputs. By addressing these questions, we aim to uncover underlying bias patterns and assess how context influences LLM recommendations.



To analyze bias in LLM-based recommendations, we introduce a metric called \textit{normalized fraction}, $F^i_a$. It measures the proportion of recommendations from genre $a$ given to group $i$ compared to all groups being considered. This is defined as:

\begin{small}
\begin{align*}
    F_a^i = \frac{\# \, \textrm{recommended items from genre a to class}\; i}{\# \, \textrm{recommended items from genre a to all considered classes}}
\end{align*}    
\end{small}

For example, let us consider a group of $30$ people, divided equally into three groups: $10$ students, $10$ musicians, and $10$ athletes. Suppose the recommendation system suggests $64$ rock songs to the students, $88$ rock songs to the musicians, and $48$ rock songs to the athletes. We can compute the normalized fraction for students as $
    F_{\textrm{rock}}^{\textrm{students}} = \dfrac{64}{64 + 88 + 48} = \frac{64}{200} = 0.32$.
Similarly, the normalized fractions for musicians and athletes are:
    $F_{\textrm{rock}}^{\textrm{musicians}} = 0.44, \;\;\textrm{and} \;\; F_{\textrm{rock}}^{\textrm{athletes}} = 0.24$.
Note that the sum of these fractions is unity (i.e., normalization).

\begin{figure*}[t]
\centering
\begin{subfigure}[t]{0.285\textwidth}
\captionsetup{justification=centering}
\resizebox{0.9\linewidth}{!}{
\begin{tikzpicture}
\begin{axis}[
width=5in,
height=3.203in,
at={(2.6in,0.852in)},
major x tick style = transparent,
ybar=2*\pgflinewidth,
bar width=20pt,
ymajorgrids,
xmajorgrids,
xlabel style={font=\color{white!15!black}, font = \Large},
ylabel style={font=\color{white!15!black}, font = \Large},
ylabel={Normalized Fraction of Genres},
symbolic x coords={{\LARGE Romance},{\LARGE Thriller},{\LARGE Sci-Fi}},
xtick = data,
scaled y ticks = false,
enlarge x limits= 0.3,
ymin=0,
ymax=1,
legend cell align=left,
legend style={at={(0.02,0.71)}, nodes={scale=1.6}, anchor=south west, legend cell align=left, align=left, draw=white!15!black}
    ]
    
    \addplot[style={fill=my1color,mark=none}]
            coordinates {({\LARGE Romance}, 0.65) ({\LARGE Thriller},0.19) ({\LARGE Sci-Fi}, 0.06)};

\addlegendentry{Gender: Female}

   \addplot[style={fill=mycolor1,mark=none}]
            coordinates {({\LARGE Romance},0.35) ({\LARGE Thriller},0.81) ({\LARGE Sci-Fi},0.94)};

\addlegendentry{Gender: Male}

\end{axis}

\end{tikzpicture}%
}
\vspace{-0.1 cm}
\caption{\footnotesize Gender bias in movie recommendations}
\label{fig:malefemale}
\end{subfigure}
\begin{subfigure}[t]{0.28\textwidth}
\captionsetup{justification=centering}
\resizebox{0.9\linewidth}{!}{
\begin{tikzpicture}
\begin{axis}[
width=5in,
height=3.203in,
at={(2.6in,0.852in)},
major x tick style = transparent,
ybar=2*\pgflinewidth,
bar width=20pt,
ymajorgrids,
xmajorgrids,
xlabel style={font=\color{white!15!black}, font = \Large},
ylabel style={font=\color{white!15!black}, font = \Large},
ylabel={Normalized Fraction of Genres},
symbolic x coords={{\LARGE Blues},{\LARGE Classical},{\LARGE Hip-hop}},
xtick = data,
scaled y ticks = false,
enlarge x limits= 0.3,
ymin=0,
ymax=1,
legend cell align=left,
legend style={at={(0.6,0.63)}, nodes={scale=1.6}, anchor=south west, legend cell align=left, align=left, draw=white!15!black}
    ]
    
    \addplot[style={fill=my1color,mark=none}]
            coordinates {({\LARGE Blues}, 0.17) ({\LARGE Classical},0.65) ({\LARGE Hip-hop}, 0.515)};

\addlegendentry{Age: 20 years}

   \addplot[style={fill=mycolor1,mark=none}]
            coordinates {({\LARGE Blues},0.29) ({\LARGE Classical},0.31) ({\LARGE Hip-hop},0.32)};

\addlegendentry{Age: 40 years}

   \addplot[style={fill=mycolor2,mark=none}]
            coordinates {({\LARGE Blues},0.54) ({\LARGE Classical},0.04) ({\LARGE Hip-hop},0.165)};

\addlegendentry{Age: 60 years}
\end{axis}

\end{tikzpicture}%
}
\vspace{-0.1 cm}
\caption{\footnotesize Age bias in song recommendations}
\label{fig:age204060}
\end{subfigure}
\begin{subfigure}[t]{0.285\textwidth}
\captionsetup{justification=centering}
\resizebox{0.9\linewidth}{!}{
\begin{tikzpicture}
\begin{axis}[
width=5in,
height=3.203in,
at={(2.6in,0.852in)},
major x tick style = transparent,
ybar=2*\pgflinewidth,
bar width=20pt,
ymajorgrids,
xmajorgrids,
xlabel style={font=\color{white!15!black}, font = \Large},
ylabel style={font=\color{white!15!black}, font = \Large},
ylabel={Normalized Fraction of Genres},
symbolic x coords={{\LARGE Bio},{\LARGE Sci-Fi},{\LARGE Fiction}},
xtick = data,
scaled y ticks = false,
enlarge x limits= 0.3,
ymin=0,
ymax=1,
legend cell align=left,
legend style={at={(0.44,0.66)}, nodes={scale=1.4}, anchor=south west, legend cell align=left, align=left, draw=white!15!black}
    ]
    
    \addplot[style={fill=my1color,mark=none}]
            coordinates {({\LARGE Bio}, 0.20) ({\LARGE Sci-Fi},0.315) ({\LARGE Fiction}, 0.43)};

\addlegendentry{Occupation: Student}

   \addplot[style={fill=mycolor1,mark=none}]
            coordinates {({\LARGE Bio},0.74) ({\LARGE Sci-Fi},0.07) ({\LARGE Fiction},0.05)};

\addlegendentry{Occupation: Comedian}

   \addplot[style={fill=mycolor2,mark=none}]
            coordinates {({\LARGE Bio},0.06) ({\LARGE Sci-Fi},0.615) ({\LARGE Fiction},0.52)};

\addlegendentry{Occupation: Writer}

\end{axis}

\end{tikzpicture}%
}
\vspace{-0.1 cm}
\caption{\footnotesize Occupation bias in book recommendations}
\label{fig:occupation}
\end{subfigure}

\label{figrq1demo}
\vspace{-0.2 cm}
\captionsetup{justification=justified}
\caption{\small Demographic Bias in the LLM-based recommendation system (for movies, songs and books) within CLG}
\vspace{-0.5 cm}
\end{figure*}
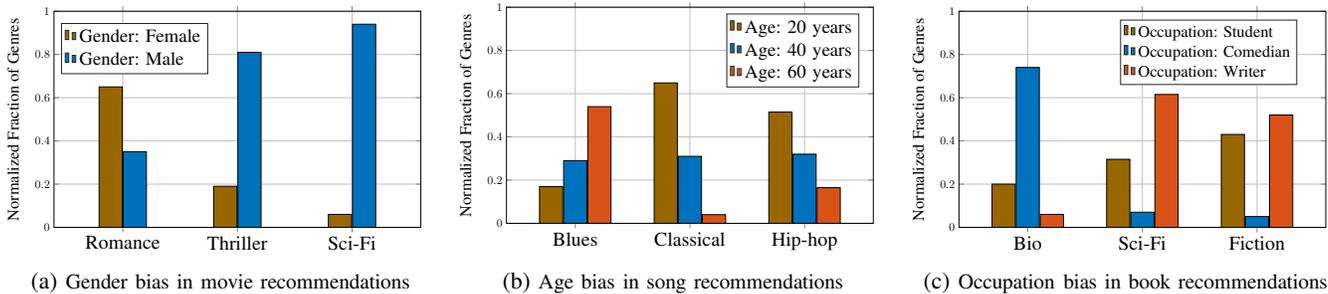
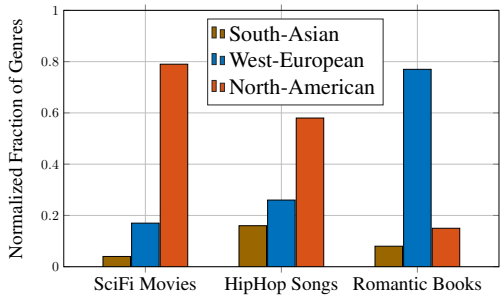
\begin{figure}[t]
\centering
\resizebox{0.65\linewidth}{!}{
\begin{tikzpicture}
\begin{axis}[
width=5in,
height=3.203in,
at={(2.6in,0.852in)},
major x tick style = transparent,
ybar=2*\pgflinewidth,
bar width=20pt,
ymajorgrids,
xmajorgrids,
xlabel style={font=\color{white!15!black}, font = \Large},
ylabel style={font=\color{white!15!black}, font = \Large},
ylabel={Normalized Fraction of Genres},
symbolic x coords={{\Large SciFi Movies},{\Large HipHop Songs},{\Large Romantic Books}},
xtick = data,
scaled y ticks = false,
enlarge x limits= 0.3,
ymin=0,
ymax=1,
legend cell align=left,
legend style={at={(0.33,0.63)}, nodes={scale=1.6}, anchor=south west, legend cell align=left, align=left, draw=white!15!black}
    ]
    
    \addplot[style={fill=my1color,mark=none}]
            coordinates {({\Large SciFi Movies}, 0.04) ({\Large HipHop Songs},0.16) ({\Large Romantic Books}, 0.08)};

\addlegendentry{South-Asian}

   \addplot[style={fill=mycolor1,mark=none}]
            coordinates {({\Large SciFi Movies},0.17) ({\Large HipHop Songs},0.26) ({\Large Romantic Books},0.77)};

\addlegendentry{West-European}

   \addplot[style={fill=mycolor2,mark=none}]
            coordinates {({\Large SciFi Movies},0.79) ({\Large HipHop Songs},0.58) ({\Large Romantic Books},0.15)};

\addlegendentry{North-American}

\end{axis}

\end{tikzpicture}%
}
\vspace{-0.07 in}
\captionsetup{justification=justified}
\caption{\footnotesize Cultural bias in LLM-based recommendations}
\label{fig:culturebias1}
\vspace{-0.1 in}
\end{figure}

\subsection{Context-less generation (CLG)}
\label{sec:clgbias}
To investigate potential biases in LLM-based recommendation systems, we start by examining recommendations generated in a context-free generation (CFG) framework. We focus on whether and how LLMs' recommendations for books, songs, and movies show demographic and cultural biases, guided by a specific set of research questions.

\vspace{0.05 in}
\noindent
\fcolorbox{black}{green!20}{
\parbox{.46\textwidth}
{{\bf RQ1}: Do certain genres of books, movies, or songs receive more frequent recommendations within the CLG?}
}
\vspace{0.05 in}



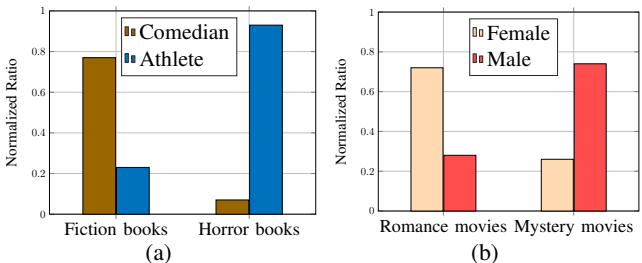
\begin{figure}[t]
\centering
\begin{subfigure}[t]{0.235\textwidth}
\captionsetup{justification=centering}
\resizebox{0.99\linewidth}{!}{
\begin{tikzpicture}
\begin{axis}[
width=4in,
height=3.203in,
at={(2.6in,0.852in)},
major x tick style = transparent,
ybar=2*\pgflinewidth,
bar width=30pt,
ymajorgrids,
xmajorgrids,
xlabel style={font=\color{white!15!black}, font = \Large},
ylabel style={font=\color{white!15!black}, font = \Large},
ylabel={Normalized Ratio},
symbolic x coords={{\LARGE Fiction books},{\LARGE Horror books}},
xtick = data,
scaled y ticks = false,
enlarge x limits= 0.5,
ymin=0,
ymax=1,
legend cell align=left,
legend style={at={(0.27,0.68)}, nodes={scale=2}, anchor=south west, legend cell align=left, align=left, draw=white!15!black}
    ]
    
    \addplot[style={fill=my1color,mark=none}]
            coordinates {({\LARGE Fiction books}, 0.77) ({\LARGE Horror books},0.07) };

\addlegendentry{Comedian}

   \addplot[style={fill=mycolor1,mark=none}]
            coordinates {({\LARGE Fiction books},0.23) ({\LARGE Horror books},0.93) };

\addlegendentry{Athlete}

\end{axis}

\end{tikzpicture}%
}
\vspace{-0.6 cm}
\caption{}
\label{fig:cbgoccupation}
\end{subfigure}
\begin{subfigure}[t]{0.23\textwidth}
\captionsetup{justification=centering}
\resizebox{0.99\linewidth}{!}{
\begin{tikzpicture}
\begin{axis}[
width=4in,
height=3.203in,
at={(2.6in,0.852in)},
major x tick style = transparent,
ybar=2*\pgflinewidth,
bar width=30pt,
ymajorgrids,
xmajorgrids,
xlabel style={font=\color{white!15!black}, font = \Large},
ylabel style={font=\color{white!15!black}, font = \Large},
ylabel={Normalized Ratio},
symbolic x coords={{\LARGE Romance movies},{\LARGE Mystery movies}},
xtick = data,
scaled y ticks = false,
enlarge x limits= 0.5,
ymin=0,
ymax=1,
legend cell align=left,
legend style={at={(0.35,0.68)}, nodes={scale=2}, anchor=south west, legend cell align=left, align=left, draw=white!15!black}
    ]
    
    \addplot[style={fill=orange!30,mark=none}]
            coordinates {({\LARGE Romance movies}, 0.72) ({\LARGE Mystery movies},0.26) };

\addlegendentry{Female}

   \addplot[style={fill=red!70,mark=none}]
            coordinates {({\LARGE Romance movies},0.28) ({\LARGE Mystery movies},0.74) };

\addlegendentry{Male}

\end{axis}

\end{tikzpicture}%
}
\vspace{-0.6 cm}
\caption{}
\label{fig:cbgagebook}
\end{subfigure}
\vspace{-0.1 in}
\captionsetup{justification=justified} 
\caption{\footnotesize Comparison of bias in LLM-based recommendation systems in LLaMA, depending on (a) fiction and horror books and (b) romance and mystery movies among different (a) occupations, and (b) genders.}
\label{llama_book_clg_rq1}
\vspace{-0.2 in}
\end{figure}

We analyze demographic and cultural biases in LLM-based recommendations within the context-less generation (CLG) framework. As shown in Figs. \ref{fig:malefemale}–\ref{fig:occupation}, notable gender, age, and occupation disparities emerge. For instance, females are offered more romance movies, whereas males receive more thriller and sci-fi suggestions. Age strongly influences music, with younger users recommended more hip-hop and older users more blues. Occupation also matters: writers are steered toward fiction, while comedians receive more biographies, potentially for material generation. Fig. \ref{fig:culturebias1} highlights cultural bias: North Americans see more sci-fi, while west Europeans receive more romantic books. Similar disparities arise in LLaMA’s outputs (Fig. \ref{llama_book_clg_rq1}), with gender-based movie and occupation-driven book recommendations. Further details can be found in \cite{sakib2024challenging}.


To delve further, we pose the following research question and address it with careful analysis.

\vspace{0.08 in}
\noindent
\fcolorbox{black}{green!20}{
\parbox{.46\textwidth}
{{\bf RQ2}: Are certain groups more likely to receive stereotypical or less diverse recommendations from the LLM in the CLG framework compared to others?}
}
\vspace{0.05 in}

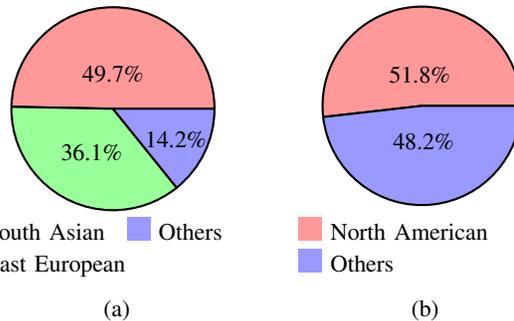
\begin{figure}[t]
\centering
\begin{subfigure}[t]{0.24\textwidth}
\centering
\captionsetup{justification=centering}
\begin{tikzpicture}
[font=\small]
    \pie[
    sum=auto,
    text=legend,
    after number=\%,
    color = {red!40, green!40, blue!40},
    radius=1.35,
    text=pin distance=1.5cm 
    ]
    {49.7/South Asian, 36.1/East European, 14.2/Others}
\end{tikzpicture}
\vspace{-0.1 in}
\begin{tabbing}
\textcolor{red!40}{\rule{0.3cm}{0.3cm}} \small South Asian \;  \textcolor{blue!40}{\rule{0.3cm}{0.3cm}} \small Others \\
\textcolor{green!40}{\rule{0.3cm}{0.3cm}} \small East European\\
\end{tabbing}
\vspace{-0.25 in}
\caption{}
\label{fig:sterclassicalsongs}
\end{subfigure}
\hspace{-0.5 cm}
\begin{subfigure}[t]{0.24\textwidth}
\centering
\captionsetup{justification=centering}
\begin{tikzpicture}
[font=\small]
    \pie[
    sum=auto,
    text=legend,
    after number=\%,
    color = {red!40, blue!40},
    radius=1.32,
    text=pin distance=1.5cm 
    ]
    {51.8/North American, 48.2/Others}
\end{tikzpicture}
\vspace{-0.1 in}
\begin{tabbing}
\;\;\;\;\;\textcolor{red!40}{\rule{0.3cm}{0.3cm}} \small North American\\
\;\;\;\;\;\textcolor{blue!40}{\rule{0.3cm}{0.3cm}}  \small Others\\
\end{tabbing}
\vspace{-0.25 in}
\caption{}
\label{ster:scifimovies}
\end{subfigure}
\captionsetup{justification=justified}
\vspace{-0.1 in}
\caption{\footnotesize (a) Classical music is highly suggested to South-Asians and East-Europeans people, and (b) SciFi movies are highly suggested to North-Americans. Note that the ``others'' category include the remaining regions.}
\label{fig:culturebias}
\end{figure}

To address this, we observe the numbers (of movies, songs or books) of recommended genres in different scenarios, and analyze potential stereotypes within different groups.
We present two examples of cultural bias in recommendation systems. First, song recommendations show a disparity: users from South Asia and Eastern Europe receive more classical music than those from other regions, as shown in Fig. \ref{fig:sterclassicalsongs}. Second, movie recommendations reveal that North American users are disproportionately suggested science fiction (SciFi) movies, as depicted in Fig. \ref{ster:scifimovies}. These findings reveal cultural stereotypes in LLM-based recommendation systems, as shown by biased content suggestions for users from different backgrounds. This suggests the algorithms perpetuate cultural biases rather than providing balanced recommendations.

Next, we state the following research question to address the impact of the bias developed by intersecting identities (e.g., occupation and gender).

\vspace{0.05 in}
\noindent
\fcolorbox{black}{green!20}{
\parbox{.46\textwidth}
{{\bf RQ3}: Do intersecting identities, (e.g., occupation and gender combined) have an additional impact on the recommendations produced by the LLM within CLG?}
}
\vspace{0.05 in}

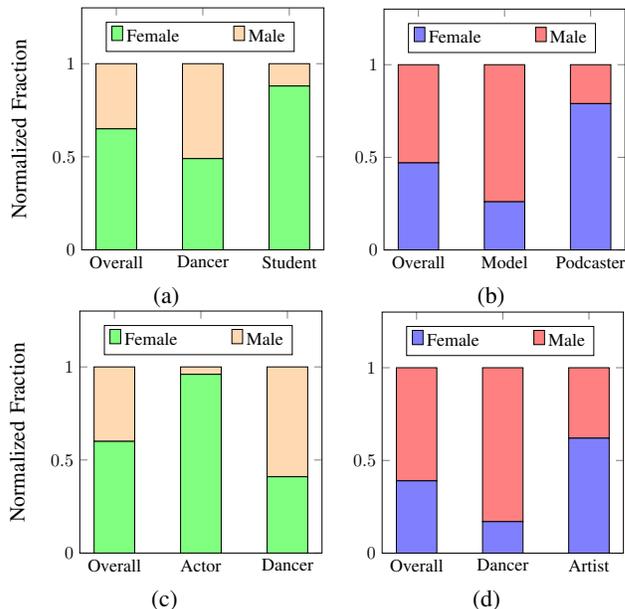
\begin{figure}[t]
\centering
\captionsetup{justification=centering}
\begin{subfigure}[t]{0.245\textwidth}
\resizebox{0.9\linewidth}{!}{
\begin{tikzpicture}
    \begin{axis}[
    width  = 2.5 in,
    height = 2.5 in,
    ybar stacked, 
    ymin=0,
    ymax=1.3,
    bar width=8mm,
    symbolic x coords={Overall,Dancer,Student},
    xtick=data,
    ylabel style={font=\color{white!15!black}, font = \large},
    ylabel={Normalized Fraction},
    enlarge x limits= 0.2,
    legend style={at={(0.1,0.88)},anchor=west,legend columns=-1,
                    /tikz/every even column/.append style={column sep=1.0cm}}
    ]

    \addplot [fill=green!50] coordinates {
        ({Overall},0.65)
        ({Dancer},0.49)
        ({Student},0.88)};
    \addplot [fill=orange!30] coordinates {
        ({Overall},0.35)
        ({Dancer},0.51)
        ({Student},0.12)};
    \legend{Female, Male}
    \end{axis}
 \end{tikzpicture} 
 }
\vspace{-0.2 cm}  
  \caption{}
\label{fig:intersectmovie}
\end{subfigure}
\begin{subfigure}[t]{0.218\textwidth}
\resizebox{0.9\linewidth}{!}{
\begin{tikzpicture}
    \begin{axis}[
    width  = 2.5 in,
    height = 2.5 in,
    ybar stacked, 
    ymin=0,
    ymax=1.3,
    bar width=8mm,
    symbolic x coords={Overall, Model, Podcaster},
    xtick=data,
    enlarge x limits= 0.2,
    legend style={at={(0.1,0.88)},anchor=west,legend columns=-1,
                    /tikz/every even column/.append style={column sep=1.0cm}}
    ]

    \addplot [fill=blue!50] coordinates {
        ({Overall},0.47)
        ({Model},0.26)
        ({Podcaster},0.79)};
    \addplot [fill=red!50] coordinates {
        ({Overall},0.53)
        ({Model},0.74)
        ({Podcaster},0.21)};
    \legend{Female, Male}
    \end{axis}
 \end{tikzpicture} 
 }
\vspace{-0.2 cm}  
 \caption{}
  \label{fig:intersectbook}
\end{subfigure}
\begin{subfigure}[t]{0.245\textwidth}
\resizebox{0.9\linewidth}{!}{
\begin{tikzpicture}
    \begin{axis}[
    width  = 2.5 in,
    height = 2.5 in,
    ybar stacked, 
    ymin=0,
    ymax=1.3,
    bar width=8mm,
    symbolic x coords={Overall,Actor,Dancer},
    xtick=data,
    ylabel style={font=\color{white!15!black}, font = \large},
    ylabel={Normalized Fraction},
    enlarge x limits= 0.2,
    legend style={at={(0.1,0.88)},anchor=west,legend columns=-1,
                    /tikz/every even column/.append style={column sep=1.0cm}}
    ]

    \addplot [fill=green!50] coordinates {
        ({Overall},0.60)
        ({Actor},0.96)
        ({Dancer},0.41)};
    \addplot [fill=orange!30] coordinates {
        ({Overall},0.40)
        ({Actor},0.04)
        ({Dancer},0.59)};
    \legend{Female, Male}
    \end{axis}
 \end{tikzpicture} 
 }
\vspace{-0.2 cm}  
  \caption{}
\label{fig:llamaintersectmovie}
\end{subfigure}
\begin{subfigure}[t]{0.214\textwidth}
\resizebox{0.9\linewidth}{!}{
\begin{tikzpicture}
    \begin{axis}[
    width  = 2.5 in,
    height = 2.5 in,
    ybar stacked, 
    ymin=0,
    ymax=1.3,
    bar width=8mm,
    symbolic x coords={Overall, Dancer, Artist},
    xtick=data,
    enlarge x limits= 0.2,
    legend style={at={(0.1,0.88)},anchor=west,legend columns=-1,
                    /tikz/every even column/.append style={column sep=1.0cm}}
    ]

    \addplot [fill=blue!50] coordinates {
        ({Overall},0.39)
        ({Dancer},0.17)
        ({Artist},0.62)};
    \addplot [fill=red!50] coordinates {
        ({Overall},0.61)
        ({Dancer},0.83)
        ({Artist},0.38)};
    \legend{Female, Male}
    \end{axis}
 \end{tikzpicture} 
 }
\vspace{-0.2 cm}  
 \caption{}
  \label{fig:llamaintersectbook}
\end{subfigure}
\vspace{-0.3 cm}
\captionsetup{justification=justified}
 \caption{\footnotesize Bias in GPT (a) romantic movie and (b) fiction book recommendations for intersecting identities (Gender-Occupation) and bias in LLaMA (c) comedy movie and (d) horror book. In particular, this shows that GPT recommends females more romantic movies than males; and LLaMA favors males in horror book suggestions compared to females.}
 \vspace{-0.2 in}
 \label{fig:rq2}
\end{figure}    

We analyzed the number of recommendations for various genres across different scenarios, observing how biases change with multiple identities. We found significant shifts in the recommendation patterns when specific identities were added.

Fig. \ref{fig:intersectmovie} illustrates the movie recommender system's bias. Generally, it suggests more romantic movies to females than males, with a normalized ratio of $0.65:0.35$. The difference has been enhanced in the case of students, where female students receive significantly more romantic movie recommendations than male students ($0.88:0.12$). However, in the case of dancers, unlike the overall trend, males and females receive similar romantic movies recommendations. Similarly, Fig. \ref{fig:intersectbook} illustrates the book recommender system's bias. Generally, it suggests fiction books to females and males quite equally. However, male models receive a significantly higher number of fiction book recommendations than female models ($0.74:0.26$). Conversely, female podcasters receive significantly more fiction book recommendations than male podcasters ($0.79:0.21$). This shows that occupation further impacts gender bias in LLM-based recommendations.

A similar pattern is observed in LLaMA, where its recommendation behavior adapts and exhibits distinct biases in the presence of intersecting identities. As shown in Figs. \ref{fig:llamaintersectmovie} and \ref{fig:llamaintersectbook}, the biases between males and females shift noticeably based on their occupations. For example, while there is an overall negligible bias in comedy movie recommendations between males and females, female actors receive substantially more comedy movie recommendations compared to male actors. Similarly, female dancers are recommended significantly fewer horror books compared to their male counterparts. These results underscore the influence of intersecting identities on LLaMA's recommendation biases.
\vspace{-0.15 in}

\subsection{Context-based generation (CBG)}

\label{sec:cbgbias}
We now analyze LLM-based recommendations within CBG (context-based generations) and investigate the \textit{impact of context} compared to CLG. To explore this, we state the following research questions and address them with examples.

\vspace{0.05 in}
\noindent
\fcolorbox{black}{green!20}{
\parbox{.46\textwidth}
{{\bf RQ4}: What is the impact on the fairness of contextual information in LLM-based recommendations when considering CBG, compared to CLG?}
}
\vspace{0.05 in}

We observe the number of genres recommended (movies, songs, books) within CBG. First, we explore occupation bias in recommending biographic books. In CLG, comedians receive more biographic book suggestions than writers (ratio $0.92:0.08$). However, with the presence of different contexts in CBG, this ratio reduces to $0.79:0.21$, as shown in Fig. \ref{fig:cbgoccupation}.

\begin{figure}[t]
\centering
\begin{subfigure}[t]{0.23\textwidth}
\captionsetup{justification=centering}
\resizebox{0.85\linewidth}{!}{
\begin{tikzpicture}
\begin{axis}[
width=4in,
height=3.203in,
at={(2.6in,0.852in)},
major x tick style = transparent,
ybar=2*\pgflinewidth,
bar width=30pt,
ymajorgrids,
xmajorgrids,
xlabel style={font=\color{white!15!black}, font = \Large},
ylabel style={font=\color{white!15!black}, font = \Large},
ylabel={Normalized Ratio},
symbolic x coords={{\LARGE CLG},{\LARGE CBG}},
xtick = data,
scaled y ticks = false,
enlarge x limits= 0.35,
ymin=0,
ymax=1,
legend cell align=left,
legend style={at={(0.22,0.68)}, nodes={scale=2}, anchor=south west, legend cell align=left, align=left, draw=white!15!black}
    ]
    
    \addplot[style={fill=my1color,mark=none}]
            coordinates {({\LARGE CLG}, 0.92) ({\LARGE CBG},0.79) };

\addlegendentry{Comedian}

   \addplot[style={fill=mycolor1,mark=none}]
            coordinates {({\LARGE CLG},0.08) ({\LARGE CBG},0.21) };

\addlegendentry{Writer}

\end{axis}

\end{tikzpicture}%
}
\vspace{-0.3 cm}
\caption{}
\label{fig:cbgoccupation}
\end{subfigure}
\begin{subfigure}[t]{0.23\textwidth}
\captionsetup{justification=centering}
\resizebox{0.85\linewidth}{!}{
\begin{tikzpicture}
\begin{axis}[
width=4in,
height=3.203in,
at={(2.6in,0.852in)},
major x tick style = transparent,
ybar=2*\pgflinewidth,
bar width=30pt,
ymajorgrids,
xmajorgrids,
xlabel style={font=\color{white!15!black}, font = \Large},
ylabel style={font=\color{white!15!black}, font = \Large},
ylabel={Normalized Ratio},
symbolic x coords={{\LARGE CLG},{\LARGE CBG}},
xtick = data,
scaled y ticks = false,
enlarge x limits= 0.35,
ymin=0,
ymax=1,
legend cell align=left,
legend style={at={(0.25,0.68)}, nodes={scale=2}, anchor=south west, legend cell align=left, align=left, draw=white!15!black}
    ]
    
    \addplot[style={fill=orange!30,mark=none}]
            coordinates {({\LARGE CLG}, 0.64) ({\LARGE CBG},0.71) };

\addlegendentry{Age: 20}

   \addplot[style={fill=red!70,mark=none}]
            coordinates {({\LARGE CLG},0.36) ({\LARGE CBG},0.29) };

\addlegendentry{Age: 60}

\end{axis}

\end{tikzpicture}%
}
\vspace{-0.3 cm}
\caption{}
\label{fig:cbgagebook}
\end{subfigure}
\begin{subfigure}[t]{0.23\textwidth}
\captionsetup{justification=centering}
\resizebox{0.85\linewidth}{!}{
\begin{tikzpicture}
\begin{axis}[
width=4in,
height=3.203in,
at={(2.6in,0.852in)},
major x tick style = transparent,
ybar=2*\pgflinewidth,
bar width=30pt,
ymajorgrids,
xmajorgrids,
xlabel style={font=\color{white!15!black}, font = \Large},
ylabel style={font=\color{white!15!black}, font = \Large},
ylabel={Normalized Ratio},
symbolic x coords={{\LARGE CLG},{\LARGE CBG}},
xtick = data,
scaled y ticks = false,
enlarge x limits= 0.45,
ymin=0,
ymax=1,
legend cell align=left,
legend style={at={(0.33,0.67)}, nodes={scale=2}, anchor=south west, legend cell align=left, align=left, draw=white!15!black}
    ]
    
    \addplot[style={fill=magenta!30,mark=none}]
            coordinates { ({\LARGE CLG},0.40) ({\LARGE CBG}, 0.20)};

\addlegendentry{Age: 20}

   \addplot[style={fill=violet,mark=none}]
            coordinates { ({\LARGE CLG},0.60) ({\LARGE CBG},0.80)};

\addlegendentry{Age: 60}
\end{axis}

\end{tikzpicture}%
}
\vspace{-0.3 cm}
\caption{}
\label{fig:cbgagesong}
\end{subfigure}
\begin{subfigure}[t]{0.23\textwidth}
\captionsetup{justification=centering}
\resizebox{0.85\linewidth}{!}{
\begin{tikzpicture}
\begin{axis}[
width=4in,
height=3.203in,
at={(2.6in,0.852in)},
major x tick style = transparent,
ybar=2*\pgflinewidth,
bar width=30pt,
ymajorgrids,
xmajorgrids,
xlabel style={font=\color{white!15!black}, font = \Large},
ylabel style={font=\color{white!15!black}, font = \Large},
ylabel={Normalized Ratio},
symbolic x coords={{\LARGE CLG},{\LARGE CBG}},
xtick = data,
scaled y ticks = false,
enlarge x limits= 0.45,
ymin=0,
ymax=1,
legend cell align=left,
legend style={at={(0.27,0.67)}, nodes={scale=2}, anchor=south west, legend cell align=left, align=left, draw=white!15!black}
    ]
    
    \addplot[style={fill=mycolor3,mark=none}]
            coordinates { ({\LARGE CLG},0.81) ({\LARGE CBG}, 0.65)};

\addlegendentry{Male}

   \addplot[style={fill=mycolor2,mark=none}]
            coordinates { ({\LARGE CLG},0.19) ({\LARGE CBG},0.35)};

\addlegendentry{Female}
\end{axis}

\end{tikzpicture}%
}
\vspace{-0.3 cm}
\caption{}
\label{fig:cbggender}
\end{subfigure}
\label{figrq1demo}
\vspace{-0.1 in}
\captionsetup{justification=justified} 
\caption{\footnotesize Comparison of bias in LLM-based recommendation systems between the CLG and CBG scenarios, depending on (a) biography-type books, (b) Sci-Fi books and (c) Jazz songs and (d) thriller movies recommendations among different (a) occupations, (b, c) ages and (d) genders.}
\vspace{-0.15 in}
\end{figure}
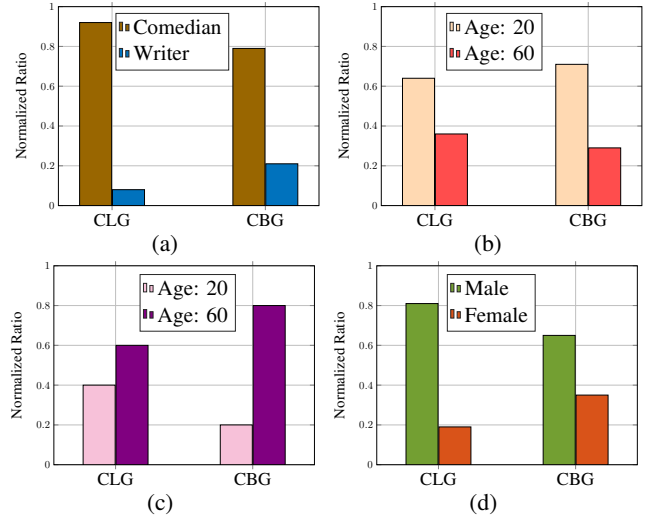

Next in Figs. \ref{fig:cbgagebook} and \ref{fig:cbgagesong} demonstrate the comparison of age bias in the recommendations for Sci-Fi books and Jazz songs, respectively. This shows that the bias could be enhanced depending on the contexts, e.g., the normalized fraction ratio of Jazz songs between 60 and 20 years old people has changed to $0.8:0.2$ within CBG, compared to $0.6:0.4$ in CLG.

Another example in Fig. \ref{fig:cbggender} shows that in CLG, LLM-based recommendations predominantly suggest thriller movies to males. However, with different contexts, more thriller movies are recommended to females.  Fig. \ref{fig:cbggender} depicts this change in the normalized fraction ratio of thriller movie recommendations to males and females.

\vspace{0.07 in}
\noindent
\fcolorbox{black}{green!20}{
\parbox{.46\textwidth}
{{\bf RQ5}: To what extent do LLM-based recommendations exhibit bias in contextual scenarios associated with CBG?}
}
\vspace{0.05 in}

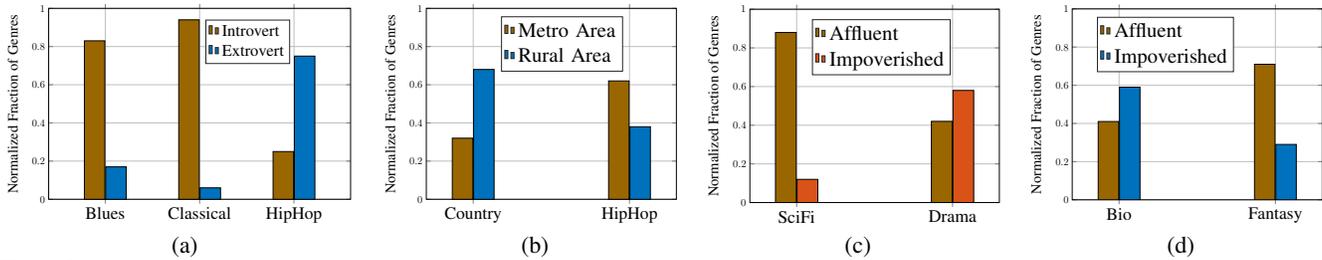
\begin{figure*}[t]
\begin{subfigure}[t]{0.27\textwidth}
\captionsetup{justification=centering}
\resizebox{0.85\linewidth}{!}{
\begin{tikzpicture}
\begin{axis}[
width=4.7in,
height=3.203in,
at={(2.6in,0.852in)},
major x tick style = transparent,
ybar=2*\pgflinewidth,
bar width=20pt,
ymajorgrids,
xmajorgrids,
xlabel style={font=\color{white!15!black}, font = \Large},
ylabel style={font=\color{white!15!black}, font = \Large},
ylabel={Normalized Fraction of Genres},
symbolic x coords={{\LARGE Blues},{\LARGE Classical},{\LARGE HipHop}},
xtick = data,
scaled y ticks = false,
enlarge x limits= 0.3,
ymin=0,
ymax=1,
legend cell align=left,
legend style={at={(0.52,0.71)}, nodes={scale=1.6}, anchor=south west, legend cell align=left, align=left, draw=white!15!black}
    ]
    
    \addplot[style={fill=my1color,mark=none}]
            coordinates {({\LARGE Blues}, 0.83) ({\LARGE Classical},0.94) ({\LARGE HipHop}, 0.25)};

\addlegendentry{Introvert}

   \addplot[style={fill=mycolor1,mark=none}]
            coordinates {({\LARGE Blues},0.17) ({\LARGE Classical},0.06) ({\LARGE HipHop},0.75)};

\addlegendentry{Extrovert}

\end{axis}

\end{tikzpicture}%
}
\vspace{-0.15 cm}
\caption{}
\label{fig:personality}
\end{subfigure}
\begin{subfigure}[t]{0.23\textwidth}
\captionsetup{justification=centering}
\resizebox{0.85\linewidth}{!}{
\begin{tikzpicture}
\begin{axis}[
width=4in,
height=3.203in,
at={(2.6in,0.852in)},
major x tick style = transparent,
ybar=2*\pgflinewidth,
bar width=20pt,
ymajorgrids,
xmajorgrids,
xlabel style={font=\color{white!15!black}, font = \Large},
ylabel style={font=\color{white!15!black}, font = \Large},
ylabel={Normalized Fraction of Genres},
symbolic x coords={{\LARGE Country},{\LARGE HipHop}},
xtick = data,
scaled y ticks = false,
enlarge x limits= 0.3,
ymin=0,
ymax=1,
legend cell align=left,
legend style={at={(0.3,0.67)}, nodes={scale=2}, anchor=south west, legend cell align=left, align=left, draw=white!15!black}
    ]
    
    \addplot[style={fill=my1color,mark=none}]
            coordinates {({\LARGE Country},0.32) ({\LARGE HipHop}, 0.62)};

\addlegendentry{Metro Area}

   \addplot[style={fill=mycolor1,mark=none}]
            coordinates {({\LARGE Country},0.68) ({\LARGE HipHop},0.38)};

\addlegendentry{Rural Area}

\end{axis}

\end{tikzpicture}%
}
\vspace{-0.15 cm}
\caption{}
\label{fig:area_first}
\end{subfigure}
\begin{subfigure}[t]{0.23\textwidth}
\captionsetup{justification=centering}
\resizebox{0.85\linewidth}{!}{
\begin{tikzpicture}
\begin{axis}[
width=3.95in,
height=3.203in,
at={(2.6in,0.852in)},
major x tick style = transparent,
ybar=2*\pgflinewidth,
bar width=20pt,
ymajorgrids,
xmajorgrids,
xlabel style={font=\color{white!15!black}, font = \Large},
ylabel style={font=\color{white!15!black}, font = \Large},
ylabel={Normalized Fraction of Genres},
symbolic x coords={{\LARGE SciFi},{\LARGE Drama}},
xtick = data,
scaled y ticks = false,
enlarge x limits= 0.3,
ymin=0,
ymax=1,
legend cell align=left,
legend style={at={(0.25,0.66)}, nodes={scale=2}, anchor=south west, legend cell align=left, align=left, draw=white!15!black}
    ]
    
    \addplot[style={fill=my1color,mark=none}]
            coordinates { ({\LARGE SciFi},0.88) ({\LARGE Drama}, 0.42)};

\addlegendentry{Affluent}

   \addplot[style={fill=mycolor2,mark=none}]
            coordinates { ({\LARGE SciFi},0.12) ({\LARGE Drama},0.58)};

\addlegendentry{Impoverished}
\end{axis}

\end{tikzpicture}%
}
\vspace{-0.15 cm}
\caption{}
\label{fig:finance}
\end{subfigure}
\begin{subfigure}[t]{0.23\textwidth}
\captionsetup{justification=centering}
\resizebox{0.85\linewidth}{!}{
\begin{tikzpicture}
\begin{axis}[
width=4in,
height=3.203in,
at={(2.6in,0.852in)},
major x tick style = transparent,
ybar=2*\pgflinewidth,
bar width=20pt,
ymajorgrids,
xmajorgrids,
xlabel style={font=\color{white!15!black}, font = \Large},
ylabel style={font=\color{white!15!black}, font = \Large},
ylabel={Normalized Fraction of Genres},
symbolic x coords={{\LARGE Bio},{\LARGE Fantasy}},
xtick = data,
scaled y ticks = false,
enlarge x limits= 0.3,
ymin=0,
ymax=1,
legend cell align=left,
legend style={at={(0.1,0.67)}, nodes={scale=2}, anchor=south west, legend cell align=left, align=left, draw=white!15!black}
    ]
    
    \addplot[style={fill=my1color,mark=none}]
            coordinates {({\LARGE Bio},0.41) ({\LARGE Fantasy}, 0.71)};

\addlegendentry{Affluent}

   \addplot[style={fill=mycolor1,mark=none}]
            coordinates {({\LARGE Bio},0.59) ({\LARGE Fantasy},0.29)};

\addlegendentry{Impoverished}

\end{axis}

\end{tikzpicture}%
}
\vspace{-0.15 cm}
\caption{}
\label{fig:area}
\end{subfigure}
\label{figrq1demo}
\vspace{-0.1 in}
\captionsetup{justification=justified} 
\caption{\footnotesize Bias in the LLM-based (a-b) song, (c) movie, and (d) book recommendation system within CBG depending on the contexts, such as (a) personality (extrovert or introvert), (b) living area (metro or rural), and (c-d) socioeconomic status (affluent or impoverished).}
\vspace{-0.2 in}
\end{figure*}

To investigate this, we analyze the numbers of recommendations in different scenario of varying contexts, and observe some interesting events. For example, the LLM-based system suggests blues or classical songs more to introverts and HipHop songs more to extroverts, indicating an obvious bias, as shown in Fig. \ref{fig:personality}. Furthermore, from Fig. \ref{fig:area_first}, we notice that HipHop songs are more recommended to the metro area people, while country songs are more recommended to the rural area people.

In addition, as we observe in Fig. \ref{fig:finance}, SciFi movies are significantly more recommended to affluent people compared to the impoverished ones, whereas dramas are more recommended to the impoverished people. Moreover, Fig. \ref{fig:area} shows that fantasy books are significantly more recommended to affluent people compared to the impoverished ones, whereas biographies are more recommended to the impoverished ones.

\begin{figure}[t]
\centering
\begin{subfigure}[t]{0.23\textwidth}
\captionsetup{justification=centering}
\resizebox{0.89\linewidth}{!}{
\begin{tikzpicture}
\begin{axis}[
width=4in,
height=3.203in,
at={(2.6in,0.852in)},
major x tick style = transparent,
ybar=2*\pgflinewidth,
bar width=30pt,
ymajorgrids,
xmajorgrids,
xlabel style={font=\color{white!15!black}, font = \Large},
ylabel style={font=\color{white!15!black}, font = \Large},
ylabel={Normalized Fraction Ratio},
symbolic x coords={{\LARGE Horror},{\LARGE Biography}},
xtick = data,
scaled y ticks = false,
enlarge x limits= 0.35,
ymin=0,
ymax=1,
legend cell align=left,
legend style={at={(0.22,0.68)}, nodes={scale=2}, anchor=south west, legend cell align=left, align=left, draw=white!15!black}
    ]
    
    \addplot[style={fill=my1color,mark=none}]
            coordinates {({\LARGE Horror}, 0.69) ({\LARGE Biography},0.39) };

\addlegendentry{Affluent}

   \addplot[style={fill=mycolor1,mark=none}]
            coordinates {({\LARGE Horror},0.31) ({\LARGE Biography},0.61) };

\addlegendentry{Impoverished}

\end{axis}

\end{tikzpicture}%
}
\vspace{-0.6 cm}
\caption{}
\label{fig:llama_personality}
\end{subfigure}
\begin{subfigure}[t]{0.23\textwidth}
\captionsetup{justification=centering}
\resizebox{0.89\linewidth}{!}{
\begin{tikzpicture}
\begin{axis}[
width=4in,
height=3.203in,
at={(2.6in,0.852in)},
major x tick style = transparent,
ybar=2*\pgflinewidth,
bar width=30pt,
ymajorgrids,
xmajorgrids,
xlabel style={font=\color{white!15!black}, font = \Large},
ylabel style={font=\color{white!15!black}, font = \Large},
ylabel={Normalized Fraction Ratio},
symbolic x coords={{\LARGE Classical},{\LARGE Hip-hop}},
xtick = data,
scaled y ticks = false,
enlarge x limits= 0.35,
ymin=0,
ymax=1,
legend cell align=left,
legend style={at={(0.25,0.68)}, nodes={scale=2}, anchor=south west, legend cell align=left, align=left, draw=white!15!black}
    ]
    
    \addplot[style={fill=orange!30,mark=none}]
            coordinates {({\LARGE Classical}, 0.80) ({\LARGE Hip-hop},0.25) };

\addlegendentry{Introvert}

   \addplot[style={fill=red!70,mark=none}]
            coordinates {({\LARGE Classical},0.20) ({\LARGE Hip-hop},0.75) };

\addlegendentry{Extrovert}

\end{axis}

\end{tikzpicture}%
}
\vspace{-0.6 cm}
\caption{}
\label{fig:llama_status}
\end{subfigure}
\vspace{-0.1 in}
\captionsetup{justification=justified} 
\caption{\footnotesize Bias in the LLaMA models on (a) books, and (b) songs recommendation system within CBG depending on the contexts: (a) socioeconomic status (affluent or impoverished), and (b) personality (extrovert or introvert).}
\label{fig:llama_cbg_new}
\vspace{-0.2 in}
\end{figure}
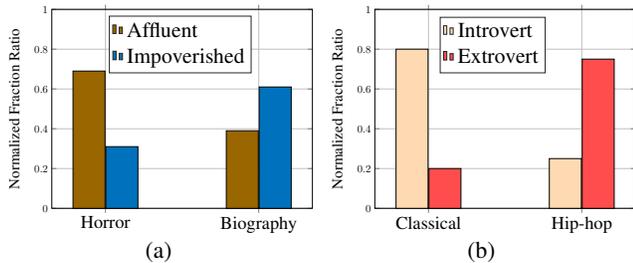
Similar trends are evident in the LLaMA model, as depicted in Fig. \ref{fig:llama_status}. This figure illustrates that horror books are predominantly recommended to affluent individuals, whereas biography books are more frequently suggested to those from impoverished backgrounds. Additionally, Fig. \ref{fig:llama_personality} highlights a distinct pattern in music preferences; classical music is more often recommended to introverts, while hip-hop is favored among extroverts. These observations underscore a significant bias in the LLM-based recommendation system, influenced by contextual factors within the Cultural Bias Group (CBG).

\vspace{0.08 in}
\noindent
\fcolorbox{black}{green!20}{
\parbox{.46\textwidth}
{{\bf RQ6}: What is the impact of the combination of contextual bias with either demographic bias or cultural bias in LLM-based recommendations?}
}
\vspace{0.07 in}

To address this question, we first observe the impact of the combination of demographic bias with the given context, and observe that there can be a significant impact of the context in terms of fairness. For instance,  in Fig. \ref{fig:gendercontext}, we demonstrate the normalized ratio for recommended number of R\&B songs and horror movies for males and females of different personality. First we observe that, the LLM recommends more R\&B songs to females compared to males. Additionally, it suggests more R\&B songs to the extroverts compared to introverts for both males or females. This shows a clear impact of an additional bias even in the presence of demographic bias. In the horror movies scenario, while we observe very little bias between female extroverts and female introverts, there exists a considerable bias between male extroverts and male introverts. Thus Fig. \ref{fig:gendercontext}  depicts scenarios which demonstrate the impact of the contexts in the presence of demographic information.

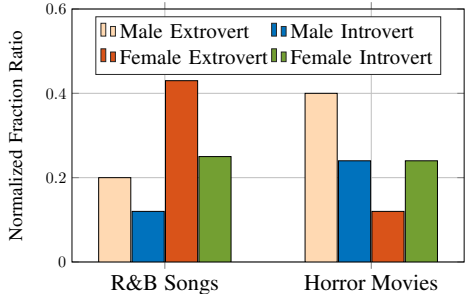
\begin{figure}[t]
\centering
\resizebox{0.7\linewidth}{!}{
\begin{tikzpicture}
\begin{axis}[
width=4in,
height=2.803in,
at={(2.6in,0.852in)},
major x tick style = transparent,
ybar=2*\pgflinewidth,
bar width=20pt,
ymajorgrids,
xmajorgrids,
xlabel style={font=\color{white!15!black}, font = \Large},
ylabel style={font=\color{white!15!black}, font = \large},
ylabel={Normalized Fraction Ratio},
symbolic x coords={{\Large R\&B Songs}, {\Large Horror Movies}},
xtick = data,
scaled y ticks = false,
enlarge x limits= 0.45,
ymin=0,
ymax=0.6,
legend cell align=left,
legend style={at={(0.06,0.74)}, nodes={scale=1.3}, anchor=south west, legend cell align=left, align=left, draw=white!15!black,legend columns=2}
    ]
    
    \addplot[style={fill=orange!30,mark=none}]
    coordinates {({\Large R\&B Songs}, 0.2) ({\Large Horror Movies},0.40) };

\addlegendentry{Male Extrovert}

    \addplot[style={fill=mycolor1,mark=none}]
    coordinates {({\Large R\&B Songs}, 0.12) ({\Large Horror Movies},0.24) };

\addlegendentry{Male Introvert}

    \addplot[style={fill=mycolor2,mark=none}]
    coordinates {({\Large R\&B Songs}, 0.43) ({\Large Horror Movies},0.12) };

\addlegendentry{Female Extrovert}

    \addplot[style={fill=mycolor3,mark=none}]
    coordinates {({\Large R\&B Songs}, 0.25) ({\Large Horror Movies},0.24) };

\addlegendentry{Female Introvert}
\end{axis}

\end{tikzpicture}%
}
\vspace{-0.1 in}
\captionsetup{justification=justified} 
\caption{\footnotesize Bias in LLM-based recommendation system showing the impact of the combination of demography and context}
\label{fig:gendercontext}
\vspace{-0.2 in}
\end{figure}
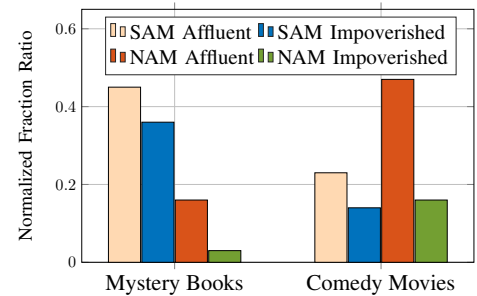
\begin{figure}[t]
\centering
\resizebox{0.65\linewidth}{!}{
\begin{tikzpicture}
\begin{axis}[
width=4in,
height=2.803in,
at={(2.6in,0.852in)},
major x tick style = transparent,
ybar=2*\pgflinewidth,
bar width=20pt,
ymajorgrids,
xmajorgrids,
xlabel style={font=\color{white!15!black}, font = \Large},
ylabel style={font=\color{white!15!black}, font = \large},
ylabel={Normalized Fraction Ratio},
symbolic x coords={{\Large Mystery Books}, {\Large Comedy Movies}},
xtick = data,
scaled y ticks = false,
enlarge x limits= 0.45,
ymin=0,
ymax=0.65,
legend cell align=left,
legend style={at={(0.06,0.74)}, nodes={scale=1.3}, anchor=south west, legend cell align=left, align=left, draw=white!15!black,legend columns=2}
    ]
    
    \addplot[style={fill=orange!30,mark=none}]
    coordinates {({\Large Mystery Books}, 0.45) ({\Large Comedy Movies},0.23) };

\addlegendentry{SAM Affluent}

    \addplot[style={fill=mycolor1,mark=none}]
    coordinates {({\Large Mystery Books}, 0.36) ({\Large Comedy Movies},0.14) };

\addlegendentry{SAM Impoverished}

    \addplot[style={fill=mycolor2,mark=none}]
    coordinates {({\Large Mystery Books}, 0.16) ({\Large Comedy Movies},0.47) };

\addlegendentry{NAM Affluent}

    \addplot[style={fill=mycolor3,mark=none}]
    coordinates {({\Large Mystery Books}, 0.03) ({\Large Comedy Movies},0.16) };

\addlegendentry{NAM Impoverished}

\end{axis}

\end{tikzpicture}%
}

\vspace{-0.2 cm}
\captionsetup{justification=justified} 
\caption{\footnotesize Bias in LLM-based recommendation system showing the impact of the combination of culture and context. In this example, we show the comparison between South-American (SAM) and North-American (NAM) people while the context includes whether they are affluent or impoverished.}
\label{fig:regioncontext}
\vspace{-0.15 in}
\end{figure}

Next in Fig. \ref{fig:regioncontext}, we demonstrate the impact of the social status context in the presence of cultural/regional information. We again consider two different scenarios which include the recommendation of mystery books and comedy movies. First, we observe that the number of recommended mystery books is always significantly lower in the impoverished class compared to the affluent people for both North-American (NAM) and South-American (SAM) people. In the comedy-movie case, while there may be a limited amount of bias between the NAM-impoverished and SAM-impoverished people, there exists a significant bias between the NAM-affluent and SAM-affluent classes. Thus, Figs. \ref{fig:gendercontext} and \ref{fig:regioncontext} motivate us to investigate the underlying bias within the recommendations due to the combination of demographic/cultural and contextual information. These findings also emphasize that context can either mitigate or intensify existing cultural and demographic biases in LLM-based recommendations.

\vspace{-0.1 in}

\subsection{Comparison among different LLMs}
\label{sec:compareLLMs}
\vspace{0.08 in}
\noindent
\fcolorbox{black}{green!20}{
\parbox{.46\textwidth}
{{\bf RQ7}: Do different Large Language Models (LLMs) vary in their performance and bias handling when applied to diverse demographic and cultural datasets?}
}
\vspace{0.07 in}

\begin{figure}[t]
\resizebox{0.9\linewidth}{!}{
\begin{tikzpicture}
\begin{axis}[
width=5in,
height=2.503in,
at={(2.6in,0.852in)},
major x tick style = transparent,
ybar=2*\pgflinewidth,
bar width=15pt,
ymajorgrids,
xmajorgrids,
xlabel style={font=\color{white!15!black}, font = \Large},
ylabel style={font=\color{white!15!black}, font = \large},
ylabel={Jensen Shannon Divergence},
symbolic x coords={{\Large Gender},{\Large Occupation},{\Large Age},{\Large Mixed}},
xtick = data,
scaled y ticks = false,
enlarge x limits= 0.15,
ymin=0,
ymax=0.5,
legend cell align=left,
legend style={at={(0.03,0.8)}, nodes={scale=1.4}, anchor=south west, legend cell align=left, align=left, draw=white!15!black, legend columns=3}
    ]
    
    \addplot[style={fill=my1color,mark=none}]
            coordinates {({\Large Gender}, 0.13) ({\Large Occupation},0.15) ({\Large Age}, 0.14)({\Large Mixed},0.41)};

\addlegendentry{GPT \,}

   \addplot[style={fill=mycolor1,mark=none}]
            coordinates {({\Large Gender},0.19) ({\Large Occupation},0.10) ({\Large Age},0.12)({\Large Mixed},0.29)};

\addlegendentry{LLaMA \,}

   \addplot[style={fill=mycolor2,mark=none}]
            coordinates {({\Large Gender},0.16) ({\Large Occupation},0.21) ({\Large Age},0.11) ({\Large Mixed},0.24)};

\addlegendentry{Gemini \,}

\end{axis}

\end{tikzpicture}%
}
\captionsetup{justification=justified}
\caption{\footnotesize Jensen–Shannon divergence for GPT, LLaMA and Gemini - recommended song genres between (a) a 20 year-female chef and a 60 year-female chef {\bf (Age)}, (b) a 40 year-female artist and a 40 year-male artist {\bf (Gender)}, (c) a 50 year-male writer and a 50 year-male comedian {\bf (Occupation)}, and (d) a 20 year-female actor and a 60 year-male athlete {\bf (Mixed)}.}
\label{fig:comparethree}
\vspace{-0.2 in}
\end{figure}
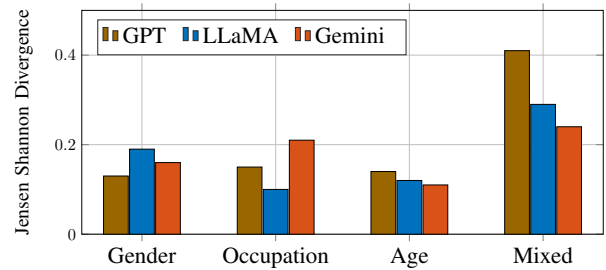

To investigate this RQ, we conduct a comparative analysis using three distinct LLMs: GPT, LLAMA, and Gemini, where we use some particular examples as mentioned in Fig. \ref{fig:comparethree}. The study utilizes the Jensen-Shannon divergence (JSD) metric to quantify the differences in recommended song genres across varied demographic categories, namely gender, occupation, age, and a combination (mixed) of all these three. The divergence values, as illustrated in the bar graph, highlight that each LLM exhibits unique patterns of response, suggesting that these models handle biases differently when processing inputs based on demographic attributes. For example, the Gemini model and the LLaMA model show pronounced divergence values in the occupation and gender categories respectively, indicating a possible inclination toward more significant bias compared to the other models in the respective categories. 

Furthermore, we observe higher JSD values in the mixed case compared to the others, which indicates a higher bias in scenarios where multiple demographic factors interact, suggesting that these LLMs may struggle to handle compound demographic characteristics effectively. These results suggest that while LLMs can perform complex tasks involving demographic data, their output can still reflect underlying biases embedded during their training phase. Therefore, addressing these biases necessitates a nuanced approach, potentially involving the retraining of models with more balanced datasets or the implementation of bias-mitigation algorithms tailored to specific demographic and cultural factors.

\section{Fairness Improving Strategies}
\label{sec:mitigatebias}
In this section, we discuss two different bias reduction strategies, including fairness aware prompt engineering \cite{clemmer2024precisedebias} and retrieval augmented generation \cite{zhang2024interactive, gao2023retrieval} (RAG)\footnotemark{}. We emphasize that our goal is not to claim novelty in bias mitigation techniques, but rather to demonstrate that even simple, well-established interventions can substantially alter fairness metrics, highlighting how deeply ingrained and sensitive bias is in LLM-based recommendation systems.

\footnotetext{While effective, the iterative interplay between the predictor and adversary in \textit{adversarial debiasing} demands significant computational resources. Hence, we did not focus on this strategy in this work and left it for future exploration.}

\vspace{-0.1 in}
\subsection{Fairness aware Prompt Engineering}
\label{sec:prompt}
Fairness-aware prompt engineering involves designing prompts in a way that minimizes potential biases in LLM responses. The technique focuses on carefully crafting input prompts to reduce the model's propensity to generate biased or skewed outputs. This can include explicitly instructing the model to prioritize fairness or neutrality when responding to queries. For instance, when generating the recommendations, prompts can be designed to explicitly instruct the model so that the recommendations are inclusive of various demographic and cultural groups. An example can be the following:

\vspace{0.07 in}
\noindent
\fcolorbox{black}{orange!20}{
\parbox{.46\textwidth}
{Ashley is a 40-year-old female chef. Can you recommend 25 movies for her? {\bf Ensure that the recommendations are inclusive of various demographic and cultural groups.}}
}
\vspace{0.07 in}

The advantage of fairness-aware prompt engineering lies in its straightforward implementation, allowing users to leverage existing models without retraining or extensive modifications. It can also be tailored dynamically, adjusting prompts to fit specific contexts or fairness goals. However, the effectiveness of this strategy depends heavily on the design of the prompts and the model’s inherent biases from its training data.

\subsection{Bias Mitigation: Retrieval-augmented Generation}
\label{sec:rag}
Contextual grounding with neutral knowledge sources was employed as a critical bias mitigation strategy to enhance the fairness and impartiality of the model's responses. By integrating retrieval-augmented generation (RAG) \cite{guu2020retrieval, lewis2020retrieval}, the model retrieves contextually relevant information from an external, curated knowledge base and incorporates this into its response generation process. This approach ensures that the responses are based on factual, unbiased data rather than the model's potentially skewed parameters.

We initiated this process by meticulously curating an extensive list comprising 200 movies, 200 songs, and 200 books that are recognized for their neutrality with respect to demographic and cultural information. This was generated by providing GPT \cite{openai2023chatgpt} with the following prompt:

\vspace{0.07 in}
\noindent
\fcolorbox{black}{orange!20}{
\parbox{.46\textwidth}
{Please provide three comprehensive lists, each containing 200 entries of movies, songs, and books, respectively. \bf{Ensure that the selections are culturally and demographically unbiased, representing a diverse range of perspectives and backgrounds.}}
}
\vspace{0.07 in}

We conducted a thorough analysis of the recommendations provided by ChatGPT to ensure that the curated list is indeed diverse and representative. By establishing this repository of impartial content, we aimed to create a foundational knowledge base that minimizes the influence of biased representations inherent in more culturally or demographically specific sources. 

Subsequently, we leveraged LangChain's \cite{langchain2023} capabilities to retrieve the top $10$ most contextually relevant entries from this curated list for each specific prompt. This retrieval process was designed to ensure that the contextual information provided to the model was both pertinent to the prompt and rooted in neutral knowledge sources. The selected entries were then synthesized to generate a coherent context, which was incorporated into the prompts presented to the model. The formulation of the prompts adhered to the following structure: ``{\textbf{Based on the following \{books/songs/movies\}: \{context\} \{prompt\}}''. By integrating these neutral knowledge sources directly into the prompting process, we provided the model with a balanced informational framework from which to generate responses.

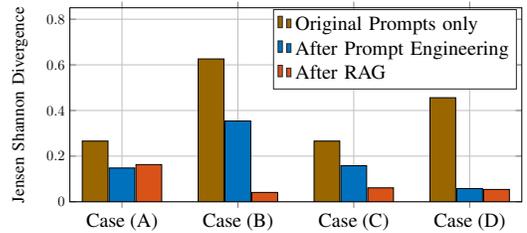
\begin{figure}[t]
\centering
\captionsetup{justification=centering}
\resizebox{0.78\linewidth}{!}{
\begin{tikzpicture}
\begin{axis}[
width=5in,
height=2.503in,
at={(2.6in,0.852in)},
major x tick style = transparent,
ybar=2*\pgflinewidth,
bar width=18pt,
ymajorgrids,
xmajorgrids,
xlabel style={font=\color{white!15!black}, font = \Large},
ylabel style={font=\color{white!15!black}, font = \large},
ylabel={Jensen Shannon Divergence},
symbolic x coords={{\Large Case (A)},{\Large Case (B)}, {\Large Case (C)},{\Large Case (D)}},
xtick = data,
scaled y ticks = false,
enlarge x limits= 0.15,
ymin=0,
ymax=0.85,
legend cell align=left,
legend style={at={(0.45,0.58)}, nodes={scale=1.4}, anchor=south west, legend cell align=left, align=left, draw=white!15!black}
    ]
    
    \addplot[style={fill=my1color,mark=none}]
            coordinates {({\Large Case (A)},0.266)({\Large Case (B)},0.626)({\Large Case (C)}, 0.266) ({\Large Case (D)},0.456) };

\addlegendentry{Original Prompts only}

   \addplot[style={fill=mycolor1,mark=none}]
            coordinates {({\Large Case (A)},0.148)({\Large Case (B)},0.354)({\Large Case (C)},0.158)  ({\Large Case (D)},0.058) };

\addlegendentry{After Prompt Engineering}

   \addplot[style={fill=mycolor2,mark=none}]
            coordinates {({\Large Case (A)},0.163)({\Large Case (B)},0.041)({\Large Case (C)},0.061)  ({\Large Case (D)},0.054) };

\addlegendentry{After RAG}

\end{axis}

\end{tikzpicture}%
}
\vspace{-0.07 in}
\captionsetup{justification=justified}
\caption{\footnotesize JSD among the distribution before and after applying mitigation methods for (A) songs for a 50-years-old male musician vs 30-years-old male actor, (B) books for a 60-years-old male musician vs 20-years-old male student, (C) movies for 60-years-old female artist vs 40-years-old female model, and (D) books for 60-years-old female actor vs 30-years-old female writer. \vspace{-4mm}}
\label{fig:promptmitigate}
\end{figure}

Next, we observe and compare the distribution of the recommendations before and after implementing the bias mitigation strategies, and measure JS-divergence between the corresponding distributions which we present in Fig. \ref{fig:promptmitigate}. We demonstrate that the JSD values have been considerably decreased after implementing the mitigation approaches which indicates a significant bias reduction in LLM-based recommendations. While we have demonstrated the JSD reduction in several scenarios within CLG, we expect that this method can also perform similarly for CBG.

\vspace{-0.1 in}
\section{Numerical Results}
\label{sec:numexp}
In this section, we discuss various bias metrics employed in fairness analysis and detail how these metrics quantify biases across several fairness-related scenarios. We explore different methodologies for measuring these biases, providing a comprehensive framework for evaluating fairness. 

\vspace{-0.1 in}
\subsection{Fairness Metrics}
We analyze three key fairness measures: statistical parity difference (SPD), disparate impact (DI), and equal opportunity difference (EOD), to quantify bias in LLM-based recommendations. SPD evaluates whether different groups receive favorable outcomes at equal rates, with a value of zero indicating perfect fairness. DI compares the ratio of favorable outcomes between groups, where a value of one reflects fairness. EOD focuses on whether individuals with positive ground-truth labels receive favorable predictions equally across groups, with zero representing fairness. These measures provide a comprehensive view of group-level disparities in model behavior  \cite{sakib2024challenging}.

\subsection{Quantifying Bias in Context-less Generations (CLG)}
\label{sec:numclg}


We begin by addressing several fairness-related questions (FQs) and apply the bias metrics discussed above to assess the demographic and cultural biases in context-less generations (CLG) by GPT-3.5. To evaluate the system's performance in addressing these FQs, we employ Random Forest (RF) \cite{chen2023comprehensive} classifier models (with a 75\%-25\% training-testing split), which deliver notably high accuracy. A summary of the questions and corresponding metric values is provided in Table \ref{tab:fairness_metrics_demo}.

\begin{table*}[t]
\footnotesize 
\centering
\begin{tabular}{|c|c|c|c|c|c|c|c|c|c|c|c|c|}
\hline
 Does LLM-based recommendation& \multicolumn{4}{c|}{ Model Performance} & \multicolumn{4}{c|}{After prompt engineering} & \multicolumn{4}{c|}{After RAG-based mitigation}  \\ \cline{2-13}
  system suggest more &  \textbf{Acc} & \textbf{SPD} & \textbf{EOD} & \textbf{DI} &  \textbf{Acc} & \textbf{SPD} & \textbf{EOD} & \textbf{DI} &  \textbf{Acc} & \textbf{SPD} & \textbf{EOD} & \textbf{DI} \\ \hline
(1) Romantic movies to & \multirow{2}{*}{75.3} & \multirow{2}{*}{0.36}& \multirow{2}{*}{0.83} & \multirow{2}{*}{0.56}& \multirow{2}{*}{51} & \multirow{2}{*}{0.02 $\thickDarkGreenDownArrow$}& \multirow{2}{*}{0.65 $\thickDarkGreenDownArrow$} & \multirow{2}{*}{0.97 $\thickDarkGreenUpArrow$}& \multirow{2}{*}{54.8} & \multirow{2}{*}{0.09 $\thickDarkGreenDownArrow$}& \multirow{2}{*}{0.36 $\thickDarkGreenDownArrow$} & \multirow{2}{*}{0.73 $\thickDarkGreenUpArrow$}\\ 
  females compared to males?  &  & &  & &  & &  & &  & &  &\\ \hline
(2) Biography books  & \multirow{2}{*}{100} & \multirow{2}{*}{1.00}& \multirow{2}{*}{1.00} & \multirow{2}{*}{0.00} & \multirow{2}{*}{98} & \multirow{2}{*}{0.96 $\thickDarkGreenDownArrow$}& \multirow{2}{*}{0.96 $\thickDarkGreenDownArrow$} & \multirow{2}{*}{0.00 $\grayThickEqual$}& \multirow{2}{*}{57.0} & \multirow{2}{*}{0.14 $\thickDarkGreenDownArrow$}& \multirow{2}{*}{0.96 $\thickDarkGreenDownArrow$} & \multirow{2}{*}{0.85 $\thickDarkGreenUpArrow$}\\ 
 to comedians than writers?  &  & &  & &  & &  & &  & &  &\\ \hline
(3) Historic fiction books to & \multirow{2}{*}{95.0} & \multirow{2}{*}{0.90}& \multirow{2}{*}{0.90} & \multirow{2}{*}{0.00}& \multirow{2}{*}{92} & \multirow{2}{*}{0.84 $\thickDarkGreenDownArrow$}& \multirow{2}{*}{0.86 $\thickDarkGreenDownArrow$} & \multirow{2}{*}{0.13 $\thickDarkGreenUpArrow$}& \multirow{2}{*}{58.0} & \multirow{2}{*}{0.16 $\thickDarkGreenDownArrow$}& \multirow{2}{*}{0.34 $\thickDarkGreenDownArrow$} & \multirow{2}{*}{0.53 $\thickDarkGreenUpArrow$}\\ 
writers than entrepreneurs?   &  & & &&  & &  & &  & &  &\\ \hline
 (4) Hip-hop songs to 20-year & \multirow{2}{*}{76.7} & \multirow{2}{*}{0.29}& \multirow{2}{*}{0.88} & \multirow{2}{*}{0.67}& \multirow{2}{*}{72.5} & \multirow{2}{*}{0.45 $\thickMaroonUpArrow$}& \multirow{2}{*}{0.86 $\thickDarkGreenDownArrow$} & \multirow{2}{*}{0.53 $\thickMaroonDownArrow$}& \multirow{2}{*}{64.2} & \multirow{2}{*}{0.28 $\thickDarkGreenDownArrow$}& \multirow{2}{*}{0.63 $\thickDarkGreenDownArrow$} & \multirow{2}{*}{0.55 $\thickMaroonDownArrow$}\\ 
 people than the 60-year ones? &  & & & &  & &  & &  & &  &\\ \hline
 (5) Ficton books to & \multirow{2}{*}{91.0} & \multirow{2}{*}{0.82}& \multirow{2}{*}{0.86} & \multirow{2}{*}{0.05}& \multirow{2}{*}{72} & \multirow{2}{*}{0.44 $\thickDarkGreenDownArrow$}& \multirow{2}{*}{0.96 $\thickMaroonUpArrow$} & \multirow{2}{*}{0.54 $\thickDarkGreenUpArrow$}& \multirow{2}{*}{60.0} & \multirow{2}{*}{0.20 $\thickDarkGreenDownArrow$}& \multirow{2}{*}{0.44 $\thickDarkGreenDownArrow$} & \multirow{2}{*}{0.55 $\thickDarkGreenUpArrow$}\\ 
  writers than comedians?  &  & &  &&  & &  & &  & &  &\\ \hline
(6) SciFi movies to North& \multirow{2}{*}{100} & \multirow{2}{*}{0.93}& \multirow{2}{*}{0.97} & \multirow{2}{*}{0.03}& \multirow{2}{*}{56.7} & \multirow{2}{*}{0.07 $\thickDarkGreenDownArrow$}& \multirow{2}{*}{0.00 $\thickDarkGreenDownArrow$} & \multirow{2}{*}{0.93 $\thickDarkGreenUpArrow$}& \multirow{2}{*}{56.7} & \multirow{2}{*}{0.03 $\thickDarkGreenDownArrow$}& \multirow{2}{*}{0.03 $\thickDarkGreenDownArrow$} & \multirow{2}{*}{0.00 $\thickMaroonDownArrow$}\\
  Americans than South-Asians?  &  & & & &  & &  & &  & &  &\\ \hline
(7) EDM songs to Oceanians & \multirow{2}{*}{86.7} & \multirow{2}{*}{0.50}& \multirow{2}{*}{1.00} & \multirow{2}{*}{0.50}& \multirow{2}{*}{66.7} & \multirow{2}{*}{0.20 $\thickDarkGreenDownArrow$}& \multirow{2}{*}{0.87 $\thickDarkGreenDownArrow$} & \multirow{2}{*}{0.77 $\thickDarkGreenUpArrow$}& \multirow{2}{*}{86.7} & \multirow{2}{*}{0.50 $\grayThickEqual$}& \multirow{2}{*}{1.00 $\grayThickEqual$} & \multirow{2}{*}{0.50 $\grayThickEqual$}\\ 
 people than East-Asians?  &  & & &&  & &  & &  & &  &\\ \hline
(8) Mystery books to East  & \multirow{2}{*}{100} & \multirow{2}{*}{0.83}& \multirow{2}{*}{0.90} & \multirow{2}{*}{0.07} & \multirow{2}{*}{80} & \multirow{2}{*}{0.37 $\thickDarkGreenDownArrow$}& \multirow{2}{*}{0.97 $\thickMaroonUpArrow$} & \multirow{2}{*}{0.62 $\thickDarkGreenUpArrow$}& \multirow{2}{*}{60.0} & \multirow{2}{*}{0.03 $\thickDarkGreenDownArrow$}& \multirow{2}{*}{0.07 $\thickDarkGreenDownArrow$} & \multirow{2}{*}{0.50 $\thickDarkGreenUpArrow$}\\ 
  Europeans than North-Africans?  &  & & &&  & &  & &  & &  &\\ \hline
   (9) Classical songs  to East Euro- & \multirow{2}{*}{76.7} & \multirow{2}{*}{0.53}& \multirow{2}{*}{0.97} & \multirow{2}{*}{0.45}& \multirow{2}{*}{73.3} & \multirow{2}{*}{0.43 $\thickDarkGreenDownArrow$}& \multirow{2}{*}{1.00 $\thickMaroonUpArrow$} & \multirow{2}{*}{0.57 $\thickDarkGreenUpArrow$}& \multirow{2}{*}{56.7} & \multirow{2}{*}{0.03 $\thickDarkGreenDownArrow$}& \multirow{2}{*}{1.00 $\thickMaroonUpArrow$} & \multirow{2}{*}{0.97 $\thickDarkGreenUpArrow$}\\ 
    -peans than Sub-Saharan Africans?  &  & & &&  & &  &&  & &  & \\ \hline
    (10) Mystery books to South Ame- & \multirow{2}{*}{100} & \multirow{2}{*}{1.00}& \multirow{2}{*}{1.00} & \multirow{2}{*}{0.00} & \multirow{2}{*}{66.7} & \multirow{2}{*}{0.20 $\thickDarkGreenDownArrow$}& \multirow{2}{*}{0.97 $\thickDarkGreenDownArrow$} & \multirow{2}{*}{0.79 $\thickDarkGreenUpArrow$}& \multirow{2}{*}{80.0} & \multirow{2}{*}{0.50 $\thickDarkGreenDownArrow$}& \multirow{2}{*}{1.00 $\grayThickEqual$} & \multirow{2}{*}{0.50 $\thickDarkGreenUpArrow$}\\ 
    -ricans than Sub-Saharan Africans?   &  & & &&  & &  & &  & &  &\\ \hline
\end{tabular}
\vspace{-0.15 cm}
\caption{\footnotesize Model performance and fairness metrics (before and after implementing bias mitigation strategies) for demographic and cultural bias to address fairness related questions for CLG. Here, $\thickDarkGreenUpArrow$ indicates that an increase in the value corresponds to an improvement in fairness (green up arrow), $\thickMaroonDownArrow$ indicates that a decrease in the value corresponds to a degradation (maroon down arrow), $\thickDarkGreenDownArrow$ indicates that a decrease in the value corresponds to an improvement in fairness (green down arrow), $\thickMaroonUpArrow$ indicates that an increase in the value corresponds to a degradation (maroon up arrow), and $\grayThickEqual$ indicates no change (blue equal sign).}
\vspace{-0.15 cm}
\label{tab:fairness_metrics_demo}
\end{table*}

While we have several FQs stated and evaluated in Table \ref{tab:fairness_metrics_demo}, we choose an example to describe the overall process and the corresponding inherent bias. For instance, we choose the FQ-5, whether the LLM-based recommendation system suggests more fiction books to writers than comedians. We are motivated to this FQ by our analysis in Fig \ref{fig:occupation}, where we observe that writers receive a significantly higher number of fiction books than comedians. 


Table \ref{tab:fairness_metrics_demo} shows that the classifier can distinguish between writers and comedians with $91\%$ accuracy based solely on the number of recommended fiction books. Furthermore, for the recommendation process to be considered fair, both bias metrics, SPD and EOD, should ideally be close to zero. However, for FQ-5, SPD and EOD are $0.82$ and $0.86$, respectively. Additionally, the DI metric, which should be \textit{one} for a perfectly fair recommendation, is $0.05$ in this case. Therefore, we conclude that the LLM-based recommendation system exhibits significant bias within this FQ.

We also observe similar trends in different other FQs which involve various genres of books, movies or songs. For example, FQ-6 addresses the question regarding recommending science fiction (Sci-Fi) movies between North American and South Asian people. The classifier can differentiate them with $100\%$ accuracy while providing SPD = $0.93$, EOD = $0.97$ and DI = $0.07$, which again indicates a significant bias. In summary, we address different fairness-related questions (FQs) in Table \ref{tab:fairness_metrics_demo} which demonstrates high performance of the models (up to $100\%$ accuracy) along with considerable bias (SPD up to $1.0$, EOD up to $1.0$, and DI to be even $0.0$).

\begin{table*}[t]
\footnotesize 
\centering
\begin{tabular}{|c|c|c|c|c|c|c|c|c|c|}
\hline
\multirow{2}{*}{Index} &  \multirow{2}{*}{Does LLM-based recommendation system suggest} & \multicolumn{4}{c|}{CLG (absence of the context)} & \multicolumn{4}{c|}{CBG (includes the contexts)} \\ \cline{3-10}
& &  \textbf{Acc} & \textbf{SPD} & \textbf{EOD} & \textbf{DI} &  \textbf{Acc} & \textbf{SPD} & \textbf{EOD} & \textbf{DI} \\ \hline
\multirow{2}{*}{\textbf{{FQ(1)}}} & More romantic movies to affluent  writers & \multirow{2}{*}{83.00} & \multirow{2}{*}{0.66}& \multirow{2}{*}{0.76} & \multirow{2}{*}{0.132}& \multirow{2}{*}{95.50} & \multirow{2}{*}{0.87 $\thickMaroonUpArrow$}& \multirow{2}{*}{0.87 $\thickMaroonUpArrow$} & \multirow{2}{*}{0.006 $\thickMaroonDownArrow$} \\ 
 & than the impoverished athletes?   &  & & && & &  & \\ \hline
 \multirow{2}{*}{\textbf{FQ(2)}} & More Pop songs to 20y extrovert & \multirow{2}{*}{65.83} & \multirow{2}{*}{0.70}& \multirow{2}{*}{1.00} & \multirow{2}{*}{0.30}& \multirow{2}{*}{83.13} & \multirow{2}{*}{0.66 $\thickDarkGreenDownArrow$}& \multirow{2}{*}{0.79 $\thickDarkGreenDownArrow$} & \multirow{2}{*}{0.17 $\thickMaroonDownArrow$} \\ 
 &  people than 60y introvert people?   &  & &  &&& &  & \\ \hline
 \multirow{2}{*}{\textbf{{FQ(3)}}} & More reggae songs to impoverished Sub-Saharan & \multirow{2}{*}{63.33} & \multirow{2}{*}{0.267}& \multirow{2}{*}{1.00} & \multirow{2}{*}{0.733}& \multirow{2}{*}{97.50} & \multirow{2}{*}{0.89 $\thickMaroonUpArrow$}& \multirow{2}{*}{0.95 $\thickDarkGreenDownArrow$} & \multirow{2}{*}{0.066 $\thickMaroonDownArrow$} \\ 
 & African people than affluent North Americans?   &  & &  &&& &  & \\ \hline
\multirow{2}{*}{\textbf{{FQ(4)}}} & More drama-type movies to North American & \multirow{2}{*}{71.67} & \multirow{2}{*}{0.433}& \multirow{2}{*}{0.87} & \multirow{2}{*}{0.50}& \multirow{2}{*}{92.50} & \multirow{2}{*}{0.77 $\thickMaroonUpArrow$}& \multirow{2}{*}{0.87 $\grayThickEqual$} & \multirow{2}{*}{0.12 $\thickMaroonDownArrow$} \\ 
 & introvert people than East-Asian extrovert ones?   &  & & && & &  & \\ \hline
\multirow{2}{*}{\textbf{{FQ(5)}}} & More fiction books to South-Asian metro  & \multirow{2}{*}{83.33} & \multirow{2}{*}{0.667}& \multirow{2}{*}{0.80} & \multirow{2}{*}{0.167}& \multirow{2}{*}{93.33} & \multirow{2}{*}{0.78 $\thickMaroonUpArrow$}& \multirow{2}{*}{0.82 $\thickMaroonUpArrow$} & \multirow{2}{*}{0.04 $\thickMaroonDownArrow$} \\ 
 & area people than the Ocenian rural ones?   &  & & && & &  & \\ \hline
\multirow{2}{*}{\textbf{{FQ(6)}}} & More biographic books to Affluent & \multirow{2}{*}{55.00} & \multirow{2}{*}{0.10}& \multirow{2}{*}{0.96} & \multirow{2}{*}{0.896}& \multirow{2}{*}{88.00} & \multirow{2}{*}{0.69 $\thickMaroonUpArrow$}& \multirow{2}{*}{0.83 $\thickDarkGreenDownArrow$} & \multirow{2}{*}{0.33 $\thickMaroonDownArrow$} \\ 
 & actors than impoverished dancers?   &  & & && & &  & \\ \hline
\multirow{2}{*}{\textbf{{FQ(7)}}} & More rock songs to rural   & \multirow{2}{*}{67.00} & \multirow{2}{*}{0.34}& \multirow{2}{*}{0.90} & \multirow{2}{*}{0.62}& \multirow{2}{*}{89.50} & \multirow{2}{*}{0.75 $\thickMaroonUpArrow$}& \multirow{2}{*}{0.86 $\thickDarkGreenDownArrow$} & \multirow{2}{*}{0.13 $\thickMaroonDownArrow$} \\ 
 & musicians than the metro area dancers?   &  & & && & &  & \\ \hline
 \multirow{2}{*}{\textbf{{FQ(8)}}} & More mystery books to South-Asian impoverished & \multirow{2}{*}{91.67} & \multirow{2}{*}{0.833}& \multirow{2}{*}{0.867} & \multirow{2}{*}{0.038}& \multirow{2}{*}{96.67} & \multirow{2}{*}{0.95 $\thickMaroonUpArrow$}& \multirow{2}{*}{0.98 $\thickMaroonUpArrow$} & \multirow{2}{*}{0.03 $\thickMaroonDownArrow$} \\ 
 & people than the South-American Affluent ones?   &  & & && & &  & \\ \hline
  \multirow{2}{*}{\textbf{{FQ(9)}}} & More pop songs to East Asian extrovert  & \multirow{2}{*}{91.67} & \multirow{2}{*}{0.833}& \multirow{2}{*}{0.833} & \multirow{2}{*}{0.00}& \multirow{2}{*}{95.83} & \multirow{2}{*}{0.91 $\thickMaroonUpArrow$}& \multirow{2}{*}{0.98 $\thickMaroonUpArrow$} & \multirow{2}{*}{0.07 $\thickDarkGreenUpArrow$} \\ 
 & people than the Sub-Saharan introvert ones?   &  & & && & &  & \\ \hline
\end{tabular}
\vspace{-0.15 cm}
\caption{\footnotesize Values of fairness metrics for different FQs in CBG, and the corresponding CLG (which do not include the contexts in the prompts). The changes in the bias metrics values depict the impact of the contexts in bias. The notation follows the same conventions as Table \ref{tab:fairness_metrics_demo}: $\thickDarkGreenUpArrow$ or $\thickDarkGreenDownArrow$ indicate an improvement in fairness (green up or down arrow) and $\thickMaroonDownArrow$ and $\thickMaroonUpArrow$ indicate a degradation (maroon up/down arrow). $\grayThickEqual$ indicates no change in terms of fairness (blue equal sign). Thus, the results presented in this table validate the degradation in fairness due to the added contexts.}
\vspace{-0.45 cm}
\label{tab:fairness_metrics_culture}
\end{table*}

\subsection{Quantifying Bias in Context-based Generations (CBG)}
\label{sec:numcbg}

Now we quantify the bias metrics in the presence of different contexts along with the demographic and cultural information. As discussed in Sec. \ref{sec:cbgproblem}, we systematically include contexts in the prompts along three different directions: whether the person is affluent or impoverished, whether the person resides in a metro or rural area, and whether the person is introvert or extrovert. 
Building on the CLG framework outlined in Section \ref{sec:numclg}, we employ Random Forest classifiers to train our models under these varied conditions. This approach allows us to not only gauge the model's performance but also to scrutinize the bias metrics more closely. The outcomes of these assessments are systematically presented in Table \ref{tab:fairness_metrics_culture}. Additionally, we integrate corresponding CLG scenarios enabling us to observe the influence of the contexts on the fairness questions (FQs) under study. This comprehensive evaluation highlights how various contexts influence model behavior and fairness.

We present several FQs in Table \ref{tab:fairness_metrics_culture} derived from our analysis. To better illustrate the concept of inherent bias within these metrics, we select a specific example from the table. For instance, we choose the FQ-(4), whether the LLM-based recommendation system suggests more drama-type movies to North-American introvert people than East Asian extrovert ones. First, we notice that the classifier can separate North-Americans and East-Asians based on only the drama-type movies with $71.67\%$ accuracy in the absence of the context in the prompts (CLG). Next, when the prompts include the contexts, the classifier can separate North-American introvert people and East Asian extrovert people with a considerably higher accuracy, $92.50\%$. A significant degradation in the bias metrics values are seen, which indicates the impact of the added contexts (related to personality) in terms of bias. For example, SPD value has been increased to $0.77$ (in CBG) from $0.43$ (in CLG), which requires to be around {\it zero} for a fair (very low or zero bias) scenario. Additionally, the DI value has been reduced to $0.12$ from $0.50$, which needs to be around {\it one} for a fair case. These results further validate the significance of the added context.

While we describe an example with FQ-(4), the other FQs also show notable context-dependent biases, with significant shifts in bias metrics from CLG to CBG cases.. For example, in FQ-(3), SPD value has increased from $0.267$ (CLG) to $0.89$ (CBG), which is caused by the presence of the context on whether the corresponding person is affluent or impoverished. 
In FQ-(1), the EOD value has increased from $0.76$ (CLG) to $0.87$ (CBG) because of the context. 
Similarly in FQ-(7), the DI value has decreased from $0.62$ (CLG) to $0.13$ (CBG) depending on the context of the region in which an individual resides. 
In general, our findings indicate that the addition of context tends to reduce fairness: this result is expected as CBG offers a finer level of granularity compared to CLG. 


\vspace{-0.1 in}
\subsection{Bias Reduction}

Now to evaluate the performance of the bias mitigation strategies, we implemented prompt engineering and retrieval-augmented generation methods. After applying these procedures, we quantify the bias in terms of the same three metrics (SPD, EOD and DI) and present the results in Table \ref{tab:fairness_metrics_demo}.

Prompt engineering involves crafting inputs to the model in a way that aims at minimizing the influence of undesirable biases. As discussed in Sec. \ref{sec:prompt}, by adjusting the phrasing and context of prompts, we can guide the model towards more equitable outputs. Our results in Table \ref{tab:fairness_metrics_demo} demonstrate that this method can significantly improve fairness metrics across various recommendation categories. For instance, in recommending romantic movies to males versus females, prompt engineering adjusted the SPD from $0.36$ to $0.02$, the EOD from $0.83$ to $0.65$, and the DI from $0.56$ to $0.97$. Similarly, for the recommendation of EDM songs to Oceanian people to East-Asian ones, we observed an enhancement in fairness, with SPD reduced from $0.50$ to $0.20$, EOD reduced from $1.00$ to $0.87$ and DI enhanced from $0.50$ to $0.77$. These results indicate a more equitable distribution of recommendations from the LLMs. Note that there exist very few scenarios where a bias metric may have been worsened after prompt engineering (for example, while recommending mystery books to East European and North African people, the EOD value got worsened from $0.90$ to $0.97$); however, the other metrics (including SPD and DI) improved significantly in those cases.

Next, we implement a bias mitigation strategy, retrieval-augmented generation (RAG) which incorporates external, unbiased data during the recommendation process, allowing the model to generate outputs based on a broader and more balanced information set. As discussed in \ref{sec:rag}, this approach proved especially effective in scenarios where the baseline biases were pronounced. For example, in the case of recommending historic fiction books to writers and entrepreneurs, the RAG method improved the bias metrics significantly: reduced SPD from $0.90$ to $0.16$, reduced EOD from $0.90$ to $0.34$ and enhanced DI from $0.00$ to $0.53$. Another example could be the case of recommending mystery books to East European and North African people, where SPD has been reduced from $0.83$ to $0.03$, EOD from $0.90$ to $0.07$ and DI has been improved from $0.07$ to $0.50$. It should be noted that, similar to the prompt engineering approach, there exist very few scenarios where a bias metric may have been worsened after prompt engineering (for example, while recommending hip-hop songs to 20-year and 60-year old people, the DI value got worsened from $0.67$ to $0.55$); however, the other metrics (including SPD and EOD) showed improvement in those cases.

\vspace{-0.1 cm}

\section{Conclusion and Future Works}

\label{sec:conclusion}
In this paper, we have identified and highlighted the persistence of demographic and cultural biases in LLM-based recommendation systems. 
By formulating and answering a set of research questions, we have uncovered insights such as how intersecting identities can exacerbate bias, and how contextual factors, such as living in a metro or impoverished area, further impact biased outcomes. 
Our analysis, guided by state-of-the-art fairness metrics, indicated significant disparities in recommendations, with extreme measures like perfect unfairness scores, up to 100\% classifier accuracy, SPD and EOD values reaching $1.0$, and DI dropping to $0.0$. 
Our findings reveal that the biases in these systems are so deeply ingrained, that even a simple intervention like prompt engineering can significantly reduce them.
We further propose a retrieval-augmented generation strategy for more effective mitigation. Numerical experiments confirm the pervasive bias and the effectiveness of these solutions.

Our work can be extended to several key areas. This study focused on binary gender groups and limited demographic factors, which may not fully capture real-world diversity. Future research could explore fairness across a broader range of racial, ethnic, and socio-economic groups. Moreover, the computational overhead associated with the RAG-approach require careful consideration. Future work should explore the scalability of these strategies and their applicability to other forms of bias. Additionally, investigating the accuracy-fairness trade-off is crucial 
to ensure effective and equitable AI systems.

\section*{Acknowledgment}
We acknowledge the use of GPT  \cite{openai2023chatgpt} for assisting in rephrasing portions of the texts in this manuscript.

\bibliographystyle{IEEEtran}
\bibliography{citations_rev}

\newpage

\pagestyle{empty}
\appendix
\section{Appendix}

\subsection{Top Contexts}
\begin{table}[t]
    \centering
    \begin{tabular}{|P{1.7cm}|P{6.3cm}|}
        \hline
        \textbf{Demo\_Feature} & \textbf{Descriptor Items} \\
        \hline
        Female Names & [Kelly, Jessica, Ashley, Emily, Alice] \\
        \hline
        Male Names & [Joseph, Ronald, Bob,  John, Thomas] \\
        \hline
        \multirow{2}{*}{Occupations}  & [Student, Entrepreneur, Actor, Artist, Comedian, Chef, \\ & Dancer, Model, Musician, Podcaster, Athlete, Writer] \\
        \hline
        Ages & [20, 30, 40, 50, 60] \\
        \hline
    \end{tabular}
    \caption{\small Descriptors for Demographic Bias Analysis}
    \label{tab:demo_feature_list}
\end{table}

\label{app:democul}
\noindent {\bf Demographic Information Descriptors}: The descriptors for demographic information are similar to those used by Wan et al. \cite{wan2023kelly}. In particular, we adopt their established set of demographic descriptors, as summarized in Table \ref{tab:demo_feature_list}, to ensure consistency with prior studies and comparability of results. These descriptors form the foundation for constructing our prompts, enabling us to systematically analyze demographic bias across different groups within the recommendation context.

\noindent {\bf Cultural Information Descriptors}: For generating the descriptors used in our cultural bias analysis, we developed a tailored approach rather than relying on existing templates. Specifically, we first curated a list of representative regions, ensuring that the coverage reflected both geographic diversity and cultural relevance. We then prompted ChatGPT to provide a set of the most prominent names associated with each region, capturing culturally salient identifiers. Finally, these names were concatenated to construct our comprehensive list of cultural descriptors. The complete set of descriptors and their regional associations are presented in Table \ref{tab:cultural_feature_list}.

\begin{table}[t!]
    \centering
     \begin{tabular}{|P{5cm}|P{2.8cm}|}
        \hline
        General Names (30) & Regions (10)\\
        \hline
         [Li Wei, Kim Yoo-jung, Sato Yuki, Aarav, Muhammad, Fahim, Nur Aisyah, Nguyen Van Anh, Putu Ayu, Luca, Emma, Sofia,  Jan, Aleksandr, Anna,  Liam, Olivia, Santiago,  Sofia, Mateo, Maria,  Oliver, Charlotte, Mia,  Mohamed, Youssef, Ahmed, Amina, Grace, John]  & [East Asia, Southeast Asia, South Asia, Western Europe, Eastern Europe, Oceania, North America, North Africa, South America, Sub-Saharan Africa] \\
        \hline
    \end{tabular}
    \caption{\small Descriptors for Cultural Bias Analysis}
    \label{tab:cultural_feature_list}
\end{table}

\label{app:context}

\begin{table}[t]
\small 
\centering
\begin{tabular}{|P{2.40cm}|P{2.53cm}|P{2.48cm}|}
\hline
 \textbf{GPT} & \textbf{LLaMA} & \textbf{Gemini} \\ \hline
 Personality Traits & Daily Routine & Local Surroundings\\ 
 Lifestyle & Living Environment & Cognitive Style\\
 Ongoing Passions & Economic Context & Personality Profile\\
 Social Condition & Personality & Moral Outlook\\
 Residential Setting & Style Preferences & Living Standard\\ \hline

\end{tabular}
\caption{\small Top five contexts suggested by the LLMs which could have significant impact on the recommendations.}
\vspace{-0.1 in}
\label{tab:top3cont}
\end{table}

In this study, we identify the top three general contexts influencing movie, song, and book recommendations using three LLMs: GPT, LLAMA, and Gemini. 
We employ the following standardized prompt across all three models, allowing us to extract and compare their contextual preferences. 

\vspace{0.15 in}
\noindent
\fcolorbox{black}{orange!20}{
\parbox{.46\textwidth}
{What are the most significant general contexts, excluding demographics and cultural background, that would help you make more accurate and personalized recommendations on books, movies and songs? List the top five contexts that you would consider most important, based on their overall influence on preferences.}
}
\vspace{0.05 in}

This prompt is designed to reveal influential elements which each model uses when generating recommendations, reflecting their underlying training data and architectural differences. By analyzing their responses, we identify the top three general contexts that shape their recommendation behavior.

The findings, as detailed in Table \ref{tab:top3cont}, reveal that different LLMs suggested varying influential contexts, with slight dissimilarities in their listings. While each model emphasized distinct contextual elements, there are some overlaps (such as personality-related factors) too. Considering these differences, and to maintain consistency throughout our study, we select three representative contexts for analysis: personality (e.g., introvert or extrovert), living area (e.g., rural or urban), and socio-economic condition (e.g., affluent or impoverished). These categories were chosen based on their recurrence across models and their broad relevance to content preferences.

\subsection{Top 10 Genre List}
\label{app:top10}
In this study, we identify the top 10 genres across books, songs, and movies by leveraging the capabilities of three different language models: GPT, LLAMA, and Gemini. This approach allows us to systematically gather genre preferences as perceived by each LLM based on their training datasets and algorithms. We utilize a standardized prompt for each model: 

\vspace{0.1 in}
\noindent
\fcolorbox{black}{orange!20}{
\parbox{.46\textwidth}
{Could you provide a list of the {\bf top 10} book/movie/song genres, based on empirical data that reflects overall interest?}
}
\vspace{0.05 in}

\begin{table*}[!t]
\normalsize 
\centering
\begin{tabular}{|c|c|P{10cm}|}
\hline
  Genre Category & LLM Name & Top 10 Genres \\ \hline
 \multirow{6}{*}{\textbf{Books}} & \multirow{2}{*}{GPT} & Mystery,  Romance, Horror,  Science Fiction (Sci-Fi), Fantasy,  Thriller, Biography, Fiction, Historical Fiction, Non-Fiction \\ \cline{2-3}
 & \multirow{2}{*}{LLaMA} & Romance, Thriller, Mystery, Science Fiction, Fantasy, Horror, Historical Fiction, Crime, Adventure, Young Adult \\ \cline{2-3}
 & \multirow{2}{*}{Gemini} & Romance, Mystery, Fantasy, Science Fiction, Thriller, Historical Fiction,
Young Adult (YA), Biography, Self-Help, Fiction.\\ \hline

 \multirow{6}{*}{\textbf{Movies}} & \multirow{2}{*}{GPT} & Drama, Documentary, Action, Horror, Fantasy, Romance, Mystery, Thriller, Comedy, Science Fiction (Sci-Fi) \\ \cline{2-3}
 & \multirow{2}{*}{LLaMA} & Adventure, Comedy, Drama, Fantasy, Horror, Romance, Sci-Fi, Thriller, Action, Mystery
\\ \cline{2-3}
 & \multirow{2}{*}{Gemini} & Action, Adventure, Comedy, Drama, Horror, Science Fiction, Thriller/Suspense, Animation, Fantasy, Romance \\ \hline 
 
 \multirow{6}{*}{\textbf{Songs}} & \multirow{2}{*}{GPT} & Hip Hop, Classical, Country, Jazz, R\&B, Blues, Reggae,  Rock, Electronic Dance Music (EDM), Pop \\ \cline{2-3}
 & \multirow{2}{*}{LLaMA} &  Hip-hop, Electronic Dance Music (EDM), Folk, Latin, Rock, Classical, R\&B,  Country, Pop, Reggae \\ \cline{2-3}
 & \multirow{2}{*}{Gemini} & Hip-Hop, Country, Electronic Dance Music (EDM), Rock, Latin Music, R\&B, K-pop, Jazz, Pop, Classical \\ \hline

\end{tabular}
\caption{\small Top 10 book, movie and song genres recommended by GPT, LLaMA and Gemini. While they have slight difference in their recommendations, to maintain consistency in this work, we conduct our study using the genres recommended by GPT.}

\label{tab:top10}
\end{table*}

The findings presented in Table \ref{tab:top10} highlight that, while slight variations exist in the genre listings produced by different models, there is a notable degree of consensus across them. GPT, LLaMA, and Gemini all recognized core genres such as Romance, Mystery, Science Fiction, and Fantasy, which indicates a shared recognition of their dominance in popular culture and dataset representation. This overlap suggests that, regardless of architectural or training differences, the models converge toward similar high-frequency categories. Such consistency underscores the robustness of these genres as key markers of cultural and consumer preferences within textual corpora.

At the same time, the differences across models are informative, pointing toward model-specific biases or variations in training data emphasis. For example, GPT and LLaMA incorporated Horror into their top listings, possibly reflecting its presence in diverse mainstream and niche datasets. In contrast, Gemini prioritized categories like Self-Help, hinting at a stronger alignment with non-fictional or utilitarian reading preferences. These distinctions emphasize that while models agree on a central cluster of genres, they also reveal unique tendencies shaped by their data sources and optimization strategies, offering valuable insights for bias analysis.

However, for the sake of consistency in our analysis and to streamline the discussion in our paper, we selected the genre recommendations provided by the GPT model as our primary data source. This decision was influenced by the broader training data and established reliability of the GPT model in generating culturally and contextually relevant responses.

\subsection{Comparison among Recommendations}
\begin{table}[t]
    \centering
    \begin{tabular}{|P{4.5cm}|P{3.1cm}|}
        \hline
        \textbf{Movies for Ashley, a 20-
year-old female student } & \textbf{Movies for Thomas, a 40-year-old male artist } \\
        \hline
        1. Lady Bird 

2. The Edge of Seventeen

3. Booksmart

4. Little Women

5. The Perks of Being a Wallflower 

6. To All the Boys I've Loved Before

7. Love, Simon

8. 10 Things I Hate About You

9. La La Land

10. About Time

11. Mean Girls 

12. Easy A 

13. Clueless

14. Pitch Perfect

15 The Devil Wears Prada

16. The Hunger Games

17. Spider-Man: Into the Spider-Verse 

18. Everything Everywhere All at Once 

19. Divergent 

20. Get Out

21. A Quiet Place

22. Knives Out

23. Gone Girl 

24. Black Swan 

25. Harry Potter and the Sorcerer’s Stone  & 1. Basquiat 

2. At Eternity's Gate 

3. Frida

4. Loving Vincent 

5. The Square  

6. Mulholland Drive

7. Holy Motors 

8. The Holy Mountain 

9. Eraserhead 

10. Enter the Void

11. In the Mood for Love 

12. The Fall 

13. Birdman

14. Blade Runner 2049 

15. Drive

16. Blue Valentine 

17. Requiem for a Dream 

18. Manchester by the Sea 

19. Her 

20. Lost in Translation

21. Fight Club 

22. Trainspotting 

23. Only Lovers Left Alive 

24. The Neon Demon 

25. Spring Breakers
\\
        \hline
    \end{tabular}
    \caption{\small List of 25 movies for Ashley, a 20-year-old female student and Thomas, a 40-year-old male artist, recommended by GPT 5.2.}
    \label{tab:movies25}
\end{table}

Table \ref{tab:movies25} illustrates the list of movies recommended by GPT 5.2 for two different persons having different demographic background. Recommendations for Ashley, a 20-year-old female student, are dominated by Romance (5) and Comedy (7) movies emphasizing coming-of-age narratives and mainstream popular cinema. In contrast, recommendations for Thomas, a 40-year-old male artist, are largely concentrated in Drama (12), with only a single Romance title. While both lists share a common core of widely recognized genres, the relative emphasis and diversity of selections indicate a bias toward recommending Romance movies more frequently to female users, highlighting how user attributes can shape recommendation tendencies.

\ifCLASSOPTIONcaptionsoff
  \newpage
\fi

\end{document}